\documentclass[letterpaper,journal,oneside,onecolumn,draft]{IEEEtran}%
\usepackage[dvips,final]{graphicx}
\usepackage{amssymb}
\usepackage{amsmath}
\usepackage{amsfonts}%
\providecommand{\U}[1]{\protect\rule{.1in}{.1in}}
\newtheorem{theorem}{Theorem}

\newtheorem{proposition}[theorem]{Proposition}

\usepackage{subfigure}
\usepackage{graphicx}
\ifx\pdfoutput\relax\let\pdfoutput=\undefined\fi
\newcount\msipdfoutput
\ifx\pdfoutput\undefined\else
\ifcase\pdfoutput\else
\ifx\paperwidth\undefined\else
\ifdim\paperheight=0pt\relax\else\pdfpageheight\paperheight\fi
\ifdim\paperwidth=0pt\relax\else\pdfpagewidth\paperwidth\fi
\fi\fi\fi
\begin{document}

\title{Target Localization Accuracy Gain in MIMO Radar Based Systems}
\author{Hana Godrich$^{\circ}$,~Alexander M. Haimovich$^{\circ}$,~and~Rick S.
Blum$^{\dagger}$\\$^{\circ}$New Jersey Institute of Technology, Newark, NJ 07102\\$^{\dagger}$Lehigh University, Bethlehem, PA 18015-3084\\$^{\circ}$[hg44,haimovich]@njit.edu, $\dagger$rblum@eecs.lehigh.edu \\\thanks{A. M. Haimovich work was supported by the U.S. Air Force Office of
Scientific Research, Agreement FA9550-06-1-0026. R. S. Blum work was supported
by the Air Force Research Laboratory under agreement No. FA9550-06-1-0041}}
\maketitle

\begin{abstract}
This paper presents an analysis of target localization accuracy, attainable by
the use of MIMO (Multiple-Input Multiple-Output) radar systems, configured
with multiple transmit and receive sensors, widely distributed over a given
area. The Cramer-Rao lower bound (CRLB) for target localization accuracy is
developed for both coherent and non-coherent processing. Coherent processing
requires a common phase reference for all transmit and receive sensors. The
CRLB is shown to be inversely proportional to the signal effective bandwidth
in the non-coherent case, but is approximately inversely proportional to the
carrier frequency in the coherent case. We further prove that optimization
over the sensors' positions lowers the CRLB by a factor equal to the product
of the number of transmitting and receiving sensors. The best linear unbiased
estimator (BLUE) is derived for the MIMO target localization problem. The
BLUE's utility is in providing a closed form localization estimate that
facilitates the analysis of the relations between sensors locations, target
location, and localization accuracy. Geometric dilution of precision (GDOP)
contours are used to map the relative performance accuracy for a given layout
of radars over a given geographic area.

\end{abstract}


\section{Introduction}

\label{Section:Introduction}

\subsection{Background and Motivation}

Research in MIMO radar has been growing as evidenced by an increasing body of
literature \cite{HaimBlum08}-\cite{LehmannAsilo06}. Generally speaking, MIMO
radar systems employ multiple antennas to transmit multiple waveforms and
engage in joint processing of the received echoes from the target. Two main
MIMO radar architectures have evolved: with colocated antennas and with
distributed antennas. MIMO\ radar with colocated antennas makes use of
waveform diversity
\cite{JianStoica07,JianStoica06,XuLiStoica06,ForsytheBliss05,Fuhrmann04},
while MIMO radar with distributed antenna takes advantage of the spatial
diversity supported by the system configuration
\cite{HaimBlum08,FishlerHaim04,FishlerHaim04(2), Fishler06}. MIMO radar
systems have been shown to offer considerable advantages over traditional
radars in various aspects of radar operation such as the detection of slow
moving targets \cite{LehmannMTD,Robey04}, the ability to identify and separate
multiple targets \cite{Tabrikian06,JianStoicaRoberts}, and in the estimation
of target parameters such as direction-of-arrival (DOA)
\cite{Robey04,Tabrikian06}, and range-based target localization
\cite{LehmannAsilo06}. In particular, \cite{LehmannAsilo06} studies target
localization with MIMO radar systems utilizing sensors distributed over a wide area.

Conventional localization techniques include time-of-arrival (TOA),
time-difference-of-arrival (TDOA), and direction-of-arrival (DOA) based
schemes. MIMO radar system with colocated antenna can perform DOA estimation
of targets in the far-field, in which case, the received signal has a planar
wavefront. In this class of systems, extensive research has focused on
waveform optimization. In \cite{JianStoicaXie06,ForsytheBliss05,Fuhrmann04}
the signal vector transmitted by a MIMO radar system is designed to minimize
the cross-correlation of the signals bounced from various targets to improve
the parameter estimation accuracy in multiple target schemes. Some of the
waveform optimization techniques suggested in \cite{JianStoicaBliss07} are
based on the Cramer-Rao lower bound (CRLB) matrix \cite{KaySSPE93}%
,\cite{Gini1}. The CRLB is known to provide a tight bound on parameter
estimation for high signal-to-noise ratio (SNR). Several design criteria are
considered, such as minimizing the trace, determinant, and the largest
eigenvalue of the CRLB matrix, concluding that minimizing the trace of the
CRLB gives a good overall performance in terms of lowering the CRLB. In
\cite{Tabrikian06(2)}, a CRLB evaluation of the achievable angular accuracy is
derived for linear arrays with orthogonal signals. The use of orthogonal
signals is shown to provide better accuracy than correlated signals. For low
SNR scenarios, the Barankin bound is derived in \cite{Tabrikian06},
demonstrating that the use of orthogonal signals results in a lower SNR
threshold for transitioning into the region of higher estimation error.

MIMO radar systems with widely spread antennas take advantage of the
geographical spread of the deployed sensors. The multiple propagation paths,
created by the transmitted waveforms and echoes from scatterers in their paths
support target localization through either direct or indirect multilateration.
With direct multilateration, the observations collected by the sensors are
jointly processed to produce the localization estimate. With indirect
multilateration, the TOAs are estimated first, and the localization is
subsequently estimated from the TOAs. The observations and processing can also
be classified as either non-coherent or coherent. The distinction between the
two modes relies on the need for mere time synchronization between the
transmitting and receiving radars in the non-coherent case, versus the need
for both time and phase synchronization in the coherent case. Note that our
coherent/non-coherent terminology is limited to the processing for
localization. Thus, a transmitted signal may have in-phase and quadrature
components, yet the localization processing is non-coherent if it utilizes
only information in the signal envelope. In the sequel, we evaluate the
performance of localization utilizing both coherent and non-coherent processing.

MIMO radar systems belongs to the class of active localization systems, where
the signal usually travels a round trip, i.e. the signal transmitted by one
sensor in a radar system is reflected by the target and measured by the same
or a different sensor. Traditional single-antenna radar systems, performing
active range-based measurements, are well known in literature \cite{Skolnik02}%
-\cite{Levanon 04}. The target range is computed from the time it takes for
the transmitted signal to get to the target plus the travelling time of the
reflected signal back to the sensor. The range estimation accuracy is directly
proportional to the mean squared error (MSE) of the time delay estimation and
is shown to be inversely proportional to the signal effective bandwidth
\cite{Skolnik02}. A first study of the localization accuracy capability of
MIMO radar systems is provided in \cite{LehmannAsilo06}, where the Fisher
information matrix (FIM) is derived for the case of orthogonal signals with
coherent processing and widely separated antennas. The CRLB is analyzed
numerically, pointing out the dependency of the accuracy on the signal carrier
frequency in the coherent case, and its reliance on the relative locations of
the target and sensors. In \cite{LehmannAsilo06}, it is observed that the CRLB
is a function of the number of transmitting and receiving sensors, however an
analytical relation is not developed. The high accuracy capability of coherent
processing is illustrated by the use of the ambiguity function (AF). Active
range-based target localization techniques are also used in multistatic radar
systems, proposed in \cite{Chernyak98}. The TOA of a signal transmitted by a
single transmit radar, reflected by the target and received at multiple
receive antennas is used in the localization process. It is observed that
increasing the number of sensors improves localization performance, yet an
exact relation is not specified. This paper addresses deficiencies in the
literature by obtaining closed-form expressions of the CRLB for both coherent
and non-coherent cases.

Geolocation techniques has been the subject of extensive research. Geolocation
belongs to the class of passive localization systems, where the signal travels
one-way. Since these passive measurement systems employ multiple sensors,
further evaluation of existing results for geolocation systems might provide
insightful for the active case. In wireless communication, passive
measurements are used by multiple base stations for localization of a
radiating mobile phone. The localization accuracy performance is evaluated in
\cite{QiGeo06}. It is shown that the localization accuracy is inversely
proportional to the signal effective bandwidth as it does in the active
localization case. Moreover, the accuracy estimation is shown to be dependent
on the sensors/base stations locations. In navigation systems, the target
makes use of time synchronized transmission from multiple Global Positioning
Systems (GPS) to establish its location. In \cite{Lee75,LevanonGDOP00}, the
relation between the transmitting sensors location and the target localization
performance is analyzed. GDOP plots are used to demonstrate the dependency of
the attainable accuracy on the location of the GPS systems with respect to the
target. In an optimal setting of the GPS systems relative to the target
position, the best achievable accuracy is shown to be inversely proportional
to the square root of the number of participating GPS. In the sequel, we apply
the GDOP metric to evaluate the localization performance of MIMO radar.

\subsection{Main Contributions}

The main contributions of this paper are:

\begin{enumerate}
\item The CRLB of the target localization estimation error is developed for
the general case of MIMO\ radar with multiple waveforms transmission. The
analytical expressions of the CRLB are derived for the case of orthogonal
waveforms with non-coherent and coherent observations. The non-coherent case
is used as benchmark for evaluating the performance of the system with
coherent observations.

\item It is shown that the CRLB expressions for both the non-coherent and
coherent cases can be factored into two terms:\ a term incorporating the
effect of bandwidth and SNR, and another term accounting for the effect of
sensor placement.

\item The CRLB of the standard deviation of the localization estimate with
non-coherent observations is shown to be inversely proportional to the signals
averaged effective bandwidth. Dramatically higher accuracy can be obtained
from processing coherent observations. In this case, the CRLB is inversely
proportional to the carrier frequency. This gain is due to the exploitation of
phase information, and is referred to as \emph{coherency gain. }

\item Formulating a convex optimization problem, it is shown that symmetric
deployment of transmitting and receiving sensors around a target is optimal
with respect to minimizing the CRLB. The closed form solution of the
optimization problem also reveals that optimally placed $M$ transmitters and
$N$ receivers reduce the CRLB on the variance of the estimate by a factor
$MN/2.$ This is referred to as the \emph{MIMO radar gain}.

\item A closed form solution is developed for the BLUE of target localization
for coherent MIMO radars. It provides a closed form solution and a
comprehensive evaluation of the performance of the estimator's MSE. This
estimator provides insight into the relation between sensors locations, target
location, and localization accuracy through the use of the GDOP metric.
Contour maps of the GDOP, presented in this paper, provide a clear
understanding of the mutual relation between a given deployment of sensors and
the achievable accuracy at various target locations.
\end{enumerate}

The rest of the paper is organized as follows: The system model is introduced
in Section \ref{Section:MIMORadarConcept}. In Section \ref{Section:CRLB}, the
CRLB is derived for the general case of multiple transmitted waveforms.
Analytical expressions are obtained for the cases of non-coherent and coherent
observations with orthogonal signals. Optimization of the CRLB as a function
of sensor location is provided in Section \ref{Section:OptimizationOverAll}.
The performance of two localization estimators is evaluated in Section
\ref{Section:TargetEst}. To establish a better understanding of the relations
between the radar geographical spread and the target location, the GDOP metric
is introduced in Section \ref{Section:GDOP}. Finally, Section
\ref{Section:Conclusions} concludes the paper.

A comment on notation: vectors are denoted by lower-case bold, while matrices
use upper-case bold letters. The superscripts \textquotedblleft
T\textquotedblright\ and \textquotedblleft H\textquotedblright\ denote the
transpose and Hermitian operators, respectively. Complex conjugate is denoted
$\left(  {}\right)  ^{\ast}$. Points in the x-y plane are denoted in
upper-case $X=\left(  x,y\right)  .$

\section{System Model}

\label{Section:MIMORadarConcept}

We consider a widely distributed MIMO radar system with $M$ transmitting
radars and $N$ receiving radars. The receiving radars may be colocated with
the transmitting ones or individually positioned. The transmitting and
receiving radars are located in a two dimensional plane $(x,y)$. The $M$
transmitters are arbitrarily located at coordinates $T_{k}=\left(
x_{tk},y_{tk}\right)  ,$ $k=1,\ldots,M$, and the $N$ receivers are similarly
arbitrarily located at coordinates $R_{\ell}=\left(  x_{r\ell},y_{r\ell
}\right)  ,$ $\ell=1,\ldots,N.$ The set of transmitted waveforms in lowpass
equivalent form is $s_{k}\left(  t\right)  ,$ $k=1,\ldots,M,$ where
$\int_{\mathcal{T}}\left\vert s_{k}\left(  t\right)  \right\vert ^{2}dt=1,$
and $\mathcal{T}$ is the common duration of all transmitted waveforms. The
power of the transmitted waveforms is normalized such that the aggregate power
transmitted by the sensors is constant, irrespective of the number of transmit
sensors. To simplify the notation, the signal power term is embedded in the
noise variance term such that the SNR at the transmitter, denoted SNR$_{t}$
and defined as the transmitted power by a sensor divided by the noise power at
a receiving sensor, is set that a desired level. Let all transmitted waveforms
be narrowband signals with individual effective bandwidth $\beta_{k}$ defined
as $\beta_{k}^{2}=\left[  \left(  \int_{W_{k}}f^{2}\left\vert S_{k}\left(
f\right)  \right\vert ^{2}df\right)  /\left(  \int_{W_{k}}\left\vert
S_{k}\left(  f\right)  \right\vert ^{2}df\right)  \right]  $, where the
integration is over the range of frequencies with non-zero signal content
$W_{k}$ \cite{Skolnik02}. We further define the signals averaged effective
bandwidth or rms bandwidth\emph{\ }as $\beta^{2}=\frac{1}{M}\sum_{k=1}%
^{M}\beta_{k}^{2}$\ and the normalized bandwidth terms as $\beta_{R_{k}}%
=\beta_{k}/\beta$. The signals are narrowband in the sense that for a carrier
frequency of $f_{c},$ the narrowband signal assumption implies $\beta_{k}%
^{2}/$\textbf{\ }$f_{c}^{2}\ll1$ and $\beta^{2}/$\textbf{\ }$f_{c}^{2}\ll
1$\textbf{. }

The target model developed here generalizes the model in \cite{Skolnik02} to a
near-field scenario and distributed sensors. In Skolnik's model
\cite{Skolnik02}, the returns of individual point scatterers have fixed
amplitude and phase, and are independent of angle. For a moving target, the
composite return fluctuates in amplitude and phase due to the relative motion
of the scatterers. When the motion is slow, and the composite target return is
assumed to be constant over the observation time, the target conforms to the
classical Swerling case I model. We now proceed to generalize this model to a
target observed by a MIMO\ radar with distributed sensors. Assume an extended
target, composed of a collection of $Q$ individual point scatterers located at
coordinates $X_{q}=\left(  x_{q},y_{q}\right)  ,$ $q=1,\ldots,Q$. The
amplitudes $\zeta_{q}$ of the point scatterers are assumed to be mutually
independent. The pathloss and phase of a signal reflected by a scatterer, when
measured with respect to a transmitted signal $s_{k}\left(  t\right)  ,$ are
functions of the path transmitter-scatterer-receiver. Let ${\tau_{\ell k}%
}\left(  X_{q}\right)  $ denote the propagation time from transmitter $k,$ to
scatterer $q,$ to receiver $\ell,$
\begin{equation}
{\tau_{\ell k}}\left(  X_{q}\right)  =\frac{1}{c}\left(  \sqrt{\left(
x_{tk}-x_{q}\right)  ^{2}+\left(  y_{tk}-y_{q}\right)  ^{2}}+\sqrt{\left(
x_{r\ell}-x_{q}\right)  ^{2}+\left(  y_{r\ell}-y_{q}\right)  }\right)
,\label{e:tau_vq}%
\end{equation}
where $c$ is the speed of light. Our signal model assumes that the sensors are
located such that variations in the signal strength due to different target to
sensor distances can be neglected, i.e., the model accounts for the effect of
the sensors/target localizations only through time delays (or phase shifts) of
the signals. The common path loss term is embedded in $\zeta_{q}.$ The
baseband representation for the signal received at sensor $\ell$ is:
\begin{equation}
r_{\ell}\left(  t\right)  =\sum_{k=1}^{M}\sum_{q=1}^{Q}\zeta_{q}\exp\left(
{-j2\pi f_{c}\tau_{\ell k}}\left(  X_{q}\right)  \right)  s_{k}\left(
t-\tau_{\ell k}\left(  X_{q}\right)  \right)  +w_{\ell}(t),\label{e:r}%
\end{equation}
where the term ${2\pi f_{c}\tau_{\ell k}}\left(  X_{q}\right)  \ $is the phase
of a signal transmitted by sensor $k,$ reflected by scatterer $q$ located at
$X_{q},$ and received by sensor $\ell.$ Phases are measured relative to a
common phase reference assumed to be available at the transmitters and
receivers. The term $w_{\ell}\left(  t\right)  $ is circularly symmetric,
zero-mean, complex Gaussian noise, spatially and temporally white with
autocorrelation function $\sigma_{w}^{2}\delta\left(  \tau\right)  $. The
noise term is set $\sigma_{w}^{2}=1/$SNR$_{t}$, where SNR$_{t}$ is measured at
the transmitter. SNR$_{t}$ is normalized such that the aggregate transmitted
power is independent of the number of transmitting sensors. The SNR at the
receiver, due to a scatterer with amplitude $\zeta_{q}$, is SNR$_{r}%
=\left\vert \zeta_{q}\right\vert ^{2}$SNR$_{t}.$ Signals reflected from the
target combine at each of the receive antennas. For example, the resultant
signal at receive antenna $\ell$ is given by
\begin{equation}
\sum_{q=1}^{Q}\zeta_{q}s_{k}\left(  t-\tau_{\ell k}\left(  X_{q}\right)
\right)  \exp\left(  {-j2\pi f_{c}\tau_{\ell k}}\left(  X_{q}\right)  \right)
\thickapprox\zeta^{\prime}s_{k}\left(  t-\tau_{\ell k}\left(  X^{\prime
}\right)  \right)  \exp\left(  {-j2\pi f_{c}\tau_{\ell k}}\left(  X^{\prime
}\right)  \right)  ,\label{e:target}%
\end{equation}
where $\zeta^{\prime}$ and $\left(  {2\pi f_{c}\tau_{\ell k}}\left(
X^{\prime}\right)  \right)  $ are respectively the amplitude and phase given
by
\begin{equation}
\zeta^{\prime}=\left[  \left(  \sum_{q=1}^{Q}\zeta_{q}\cos\left(  {2\pi
f_{c}\tau_{\ell k}}\left(  X_{q}\right)  \right)  \right)  ^{2}+\left(
\sum_{q=1}^{Q}\zeta_{q}\sin\left(  {2\pi f_{c}\tau_{\ell k}}\left(
X_{q}\right)  \right)  \right)  ^{2}\right]  ^{1/2},
\end{equation}
and
\begin{equation}
{2\pi f_{c}\tau_{\ell k}}\left(  X^{\prime}\right)  =\tan^{-1}\frac{\sum
_{q=1}^{Q}\zeta_{q}\sin\left(  {2\pi f_{c}\tau_{\ell k}}\left(  X_{q}\right)
\right)  }{\sum_{q=1}^{Q}\zeta_{q}\cos\left(  {2\pi f_{c}\tau_{\ell k}}\left(
X_{q}\right)  \right)  }.
\end{equation}
In obtaining (\ref{e:target}), we invoked the narrowband assumption
$s_{k}\left(  t-\tau_{\ell k}\left(  X_{q}\right)  \right)  \thickapprox
s_{k}\left(  t-\tau_{\ell k}\left(  X^{\prime}\right)  \right)  $, for all
scatterers, namely that the change in the lowpass equivalent signals across
the target is negligible. It follows from this discussion that the extended
target is represented by a point scatterer of amplitude $\zeta^{\prime}$ and
time delays ${\tau_{\ell k}}\left(  X^{\prime}\right)  ,$ where all the
quantities are unknown.

While this target model is completely adequate for our needs, it is possible
to extend it slightly, at little cost. Assume a constant time offset error
$\Delta\tau$ at the receivers. Further, assume that the error is small such
that it does not impact the signal envelope, but it does impact the phase.
Then we can write the time delays $\tau_{\ell k}\left(  X^{\prime}\right)
=\tau_{\ell k}\left(  X\right)  +\Delta\tau$ for some location $X=\left(
x,y\right)  .$ The target model (\ref{e:target}) can now be expressed%
\begin{equation}
\zeta^{\prime}s_{k}\left(  t-\tau_{\ell k}\left(  X^{\prime}\right)  \right)
\exp\left(  {-j2\pi f_{c}\tau_{\ell k}}\left(  X^{\prime}\right)  \right)
\thickapprox\zeta s_{k}\left(  t-\tau_{\ell k}\left(  X\right)  \right)
\exp\left(  {-j2\pi f_{c}\tau_{\ell k}}\left(  X\right)  \right)  ,
\end{equation}
where $\zeta=\zeta^{\prime}e^{-j2\pi f_{c}\Delta\tau}$ and the narrowband
assumption was invoked once more. The composite target of (\ref{e:target}) is
then equivalent to a point scatterer of complex amplitude $\zeta$ and time
delays ${\tau_{\ell k}}\left(  X\right)  .$ For simplicity, the following
notation is used: ${\tau_{\ell k}=\tau_{\ell k}}\left(  X\right)  $. The
signal model (\ref{e:r}) becomes
\begin{equation}
r_{\ell}\left(  t\right)  =\sum_{k=1}^{M}\zeta\exp\left(  {-j2\pi f_{c}%
\tau_{\ell k}}\right)  s_{k}\left(  t-\tau_{\ell k}\right)  +w_{\ell
}(t).\label{e:r1}%
\end{equation}
We define the vector of received signals as $\mathbf{r=}\left[  r_{1}%
,r_{2},...,r_{N}\right]  ^{T}$ for later use. The radar system's goal is to
estimate the target location $X=\left(  x,y\right)  .$ The target location can
be estimated directly, for example by formulating the maximum likelihood
estimate (MLE) associated with (\ref{e:r1}). Alternatively, an indirect method
is to estimate first the time delays ${\tau_{\ell k}}.$ Subsequently, the
target location can be computed from the solution to a set of equations of the
form (\ref{e:tau_vq}), viz.,
\begin{equation}
{\tau_{\ell k}}=\frac{1}{c}\left(  \sqrt{\left(  x_{tk}-x\right)  ^{2}+\left(
y_{tk}-y\right)  ^{2}}+\sqrt{\left(  x_{r\ell}-x\right)  ^{2}+\left(
y_{r\ell}-y\right)  ^{2}}\right)  .\label{e:td}%
\end{equation}
The unknown complex amplitude $\zeta$ is treated as a nuisance parameter in
the estimation problem.

Let the unknown target location $X=\left(  x,y\right)  ,$ unknown time delays
delays $\tau_{\ell k}$, and unknown target complex amplitude $\zeta=\zeta
^{R}+j\zeta^{I},$ where the notation specifies the real and imaginary
components of $\zeta.$

We refer to the processing for estimating the target location as
\emph{non-coherent} or \emph{coherent}. The received signal introduced in
(\ref{e:r1}) is adequate for the coherent case, where the transmitting and
receiving radars are assumed to be both time and phase synchronized. As such,
the time delays information, $\tau_{\ell k}$, embedded in the phase terms may
be exploited in the estimation process by matching both amplitude and phase at
the receiver end. In contrast, non-coherent processing estimates the time
delays $\tau_{\ell k}$ from variations in the envelope of the transmitted
signals $s_{k}\left(  t\right)  .$ A common time reference is required for all
the sensors in the system. In this case, the transmitting radars are not phase
synchronized and therefore the received signal model is of the form:%

\begin{equation}
r_{\ell}\left(  t\right)  =\sum_{k=1}^{M}\alpha_{\ell k}s_{k}\left(
t-\tau_{\ell k}\right)  +w_{\ell}(t),\label{e:r1nc}%
\end{equation}
where the complex amplitude terms $\alpha_{\ell k}$ integrate the effect of
the phase offsets between the transmitting and receiving sources and the
target impact on the phase and amplitude of the transmitted signals. These
elements are treated as unknown complex amplitudes, where $\alpha_{\ell
k}=\alpha_{\ell k}^{R}+j\alpha_{\ell k}^{I}$. We define the following vector
notations:%
\begin{align}
&
\begin{array}
[c]{c}%
\mathbf{\alpha}=[\alpha_{11},\alpha_{12},...,\alpha_{\ell k},...,\alpha
_{MN}]^{T},
\end{array}
\label{eq:alpha_nc}\\
&
\begin{array}
[c]{cc}%
\mathbf{\alpha}^{R}=\operatorname{Re}\left(  \mathbf{\alpha}\right)  ; &
\mathbf{\alpha}^{I}=\operatorname{Im}\left(  \mathbf{\alpha}\right)  ,
\end{array}
\ \nonumber
\end{align}
where $\operatorname{Re}\left(  \mathbf{\cdot}\right)  $ and
$\operatorname{Im}\left(  \mathbf{\cdot}\right)  $ denote the real and
imaginary parts of a complex-valued vector/matrix. \textbf{\ }

\section{\ Localization CRLB}

\label{Section:CRLB}

The CRLB provides a lower bound for the MSE of any unbiased estimator for an
unknown parameter(s). Given a vector parameter $\mathbf{\theta},$ constituted
of elements $\theta_{i},$ the unbiased estimate $\widehat{\theta}_{i}$
satisfies the following inequality \cite{KaySSPE93}:%

\begin{equation}
\operatorname{var}\left(  \widehat{\theta}_{i}\right)  \geq\left[
\mathbf{J}^{-1}\left(  \mathbf{\theta}\right)  \right]  _{ii},\text{
\ \ }i=1,2,...\ \text{\ }\label{eq:CRLBbasic}%
\end{equation}
where $\left[  \mathbf{J}^{-1}\left(  \mathbf{\theta}\right)  \right]  _{ii}$
are the diagonal elements of the Fisher Information matrix (FIM)
$\mathbf{J}\left(  \mathbf{\theta}\right)  $. The FIM is given by:%

\begin{equation}
\mathbf{J}\left(  \mathbf{\theta}\right)  =E_{\mathbf{\theta}}\left[
\frac{\partial}{\partial\mathbf{\theta}}\log p\left(  \mathbf{r|\theta
}\right)  \left(  \frac{\partial}{\partial\mathbf{\theta}}\log p\left(
\mathbf{r|\theta}\right)  \right)  ^{T}\right]  ,\label{eq:FIMdef}%
\end{equation}
where $p\left(  \mathbf{r|\theta}\right)  $ is the joint probability density
function (pdf) of $\mathbf{r}$ conditioned on $\mathbf{\theta}$.

The CRLB is then defined:%
\begin{equation}
\mathbf{C}_{CRLB}=\left[  \mathbf{J}\left(  \mathbf{\theta}\right)  \right]
^{-1}.\label{eq:CRLB_FIM}%
\end{equation}
Sometime, it is easier to compute the FIM with respect to another vector
$\mathbf{\psi,}$ and apply the chain rule to derive the original
$\mathbf{J}\left(  \mathbf{\theta}\right)  .$ In our case, since the received
signals in both (\ref{e:r1}) and (\ref{e:r1nc}) are functions of the time
delays, $\tau_{\ell k},$ and the complex amplitudes, by the chain rule,
$\mathbf{J}\left(  \mathbf{\theta}\right)  $ can be expressed in the
alternative form \cite{KaySSPE93}:%
\begin{equation}
\mathbf{J}\left(  \mathbf{\theta}\right)  =\mathbf{PJ}\left(  \mathbf{\psi
}\right)  \mathbf{P}^{T},\label{eq:chainRule}%
\end{equation}
where\ $\mathbf{\psi}$ is a vector of unknown parameters, and it incorporates
the time delays. Matrix $\mathbf{J}\left(  \mathbf{\psi}\right)  $ is the FIM
with respect to $\mathbf{\psi,}$ and matrix $\mathbf{P}$ is the Jacobian:\
\begin{equation}
\mathbf{P}{\small =\frac{\partial\mathbf{\psi}}{\partial\mathbf{\theta}}%
.}\label{eq:P_def}%
\end{equation}

From this point onward, we develop the CRLB for the case of non-coherent and
coherent processing, separately.

\subsection{Non-coherent Processing CRLB}

\label{Section:CRLBnonCoherent}

For non-coherent Processing, there is no common phase reference among the
sensors. Consequently, the complex-valued terms $\alpha_{lk}$ incorporate
phase offsets among sensors and the effect of the target on the phase and
complex amplitude, following the definitions in (\ref{eq:alpha_nc}). The
vectors of unknown parameters is defined:%

\begin{equation}
\mathbf{\theta}_{nc}=\left[  x,y,\mathbf{\alpha}^{R}\mathbf{,\alpha}%
^{I}\right]  ^{T}.\label{eq:tetha_nc}%
\end{equation}
The process of localization by non-coherent processing depends on time delay
estimation of the signals observed at the receive sensors and also on the
location of the sensors. To gain insight into how each of the factors affects
the performance of localization, we utilize the form of the FIM given in
(\ref{eq:chainRule}). We define the vector of unknown parameters:
\begin{equation}
\mathbf{\psi}_{nc}=\left[  \mathbf{\tau},\mathbf{\alpha}^{R}\mathbf{,\alpha
}^{I}\right]  ^{T},\label{eq:psi_nc}%
\end{equation}
where $\mathbf{\alpha}$\ is given in (\ref{eq:alpha_nc}) and $\mathbf{\tau=}$
$\left[  \tau_{11},\tau_{12},...,\tau_{\ell k},...,\tau_{MN}\right]  ^{T}$. We
are interested only in the estimation of $x$ and $y$, while $\mathbf{\alpha
}^{R}\mathbf{,}$ $\mathbf{\alpha}^{I}$ act as nuisance parameters in the
estimation problem.

Given a set of known transmitted waveforms $s_{k}\left(  t-\tau_{\ell
k}\right)  $ parameterized by the unknown time delays $\tau_{\ell k},$ which
in turn are a function of the unknown target location $X=\left(  x,y\right)
$, the conditional, joint pdf of the observations at the receive sensors,
given by (\ref{e:r1nc}), is then:%

\begin{equation}
p\left(  \mathbf{r}|\mathbf{\psi}_{nc}\right)  \varpropto\exp\left\{
-\frac{1}{\sigma_{w}^{2}}\overset{N}{\underset{\ell=1}{\sum}}\int
\limits_{T}\left\vert r_{\ell}(t)-\overset{M}{\underset{k=1}{\sum}}%
\alpha_{\ell k}s_{k}\left(  t-\tau_{\ell k}\right)  \right\vert ^{2}%
dt\right\}  .\label{eq:pdf_nc}%
\end{equation}

The matrix $\mathbf{P}_{nc}$ for (\ref{eq:tetha_nc})\ and (\ref{eq:psi_nc}),
to be used in (\ref{eq:chainRule}),\ is defined as:%

\begin{equation}
\mathbf{P}_{nc}=\frac{\partial\mathbf{\psi}_{nc}}{\partial\mathbf{\theta}%
_{nc}}=\left[
\begin{array}
[c]{ccc}%
\frac{\partial}{\partial x}\mathbf{\tau}^{T} & \frac{\partial}{\partial
x}\left(  \mathbf{\alpha}^{R}\right)  ^{T} & \frac{\partial}{\partial
x}\left(  \mathbf{\alpha}^{I}\right)  ^{T}\\
\frac{\partial}{\partial y}\mathbf{\tau}^{T} & \frac{\partial}{\partial
y}\left(  \mathbf{\alpha}^{R}\right)  ^{T} & \frac{\partial}{\partial
y}\left(  \mathbf{\alpha}^{I}\right)  ^{T}\\
\frac{\partial\mathbf{\tau}}{\partial\mathbf{\alpha}^{R}} & \frac
{\partial\mathbf{\alpha}^{R}}{\partial\mathbf{\alpha}^{R}} & \frac
{\partial\mathbf{\alpha}^{I}}{\partial\mathbf{\alpha}^{R}}\\
\frac{\partial\mathbf{\tau}}{\partial\mathbf{\alpha}^{I}} & \frac
{\partial\mathbf{\alpha}^{R}}{\partial\mathbf{\alpha}^{I}} & \frac
{\partial\mathbf{\alpha}^{I}}{\partial\mathbf{\alpha}^{I}}%
\end{array}
\right]  _{\left(  2MN+2\right)  \times3MN}{\small ,}\label{eq:P_nc}%
\end{equation}
where $\frac{\partial}{\partial x}\mathbf{\tau}$ is standard notation for
taking the derivative with respect to $x$ of each element of $\mathbf{\tau,} $
and $\dfrac{\partial\mathbf{\tau}}{\partial\mathbf{\alpha}^{R}}$ denotes the
Jacobian of the vector $\mathbf{\tau}$ with respect to the vector
$\mathbf{\alpha}^{R}.$ The subscript denotes the matrix dimensions.

\begin{figure}[htp]
\centering\includegraphics[width=14 cm,height=15 cm]{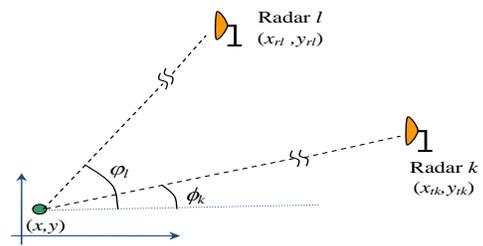}
\caption{MIMO radar system layout.}\label{Fig:1}
\end{figure}

It is not too difficult to show that using (\ref{e:td}), the matrix
$\mathbf{P}_{nc}$ can be expressed in the form:%

\begin{equation}
\mathbf{P}_{nc}={\small -}\frac{1}{c}\left[
\begin{array}
[c]{cc}%
\mathbf{H}_{2\times MN} & \mathbf{0}_{2\times2MN}\\
\mathbf{0}_{2MN\times MN} & \mathbf{I}_{2MN\times2MN}%
\end{array}
\right]  {\small ,}\label{eq:Pvalue}%
\end{equation}
where $\mathbf{0}$ is the all zero matrix, $\mathbf{I}$ is the identity
matrix, and $\mathbf{H}\in R^{2\times MN}$ incorporates the derivatives of the
time delays in (\ref{e:td}) with respect to the $x$ and $y$ parameters. These
derivatives result in cosine and sine functions of the angles the transmitting
and receiving radars create with respect to the target, incorporating
information on the sensors and target locations as follows:%

\begin{equation}
\mathbf{H}=\left[
\begin{array}
[c]{cccc}%
a_{tx_{1}}+a_{rx_{1}} & a_{tx_{1}}+a_{rx_{2}} & ... & a_{tx_{M}}+a_{rxN}\\
b_{tx_{1}}+b_{rx_{1}} & b_{tx_{1}}+b_{rx_{2}} & ... & b_{tx_{M}}+b_{rxN}%
\end{array}
\right]  .\label{eq:Hdef}%
\end{equation}
The elements of $\mathbf{H}$ are given by:%

\begin{align}
&
\begin{array}
[c]{ccc}%
a_{tx_{k}}=\cos\phi_{k}; & b_{tx_{k}}=\sin\phi_{k}; & k=1,..,M,\\
a_{rx_{\ell}}=\cos\varphi_{\ell}; & b_{rx_{\ell}}=\sin\varphi_{\ell}; &
\ell=1,..,N,
\end{array}
\label{eq:abDef}\\
&
\begin{array}
[c]{cc}%
\phi_{k}=\tan^{-1}\left(  \frac{y-y_{tk}}{x-x_{tk}}\right)  ; & \varphi_{\ell
}=\tan^{-1}\left(  \frac{y-y_{r\ell}}{x-x_{r\ell}}\right)  ,
\end{array}
\nonumber
\end{align}
where the phase $\phi_{k}$ is the bearing angle of the transmitting sensor $k$
to the target measured with respect to the $x$ axis; the phase $\varphi_{\ell
}$ is the bearing angle of the receiving radar $\ell$ to the target measured
with respect to the $x$ axis. See illustration in Figure \ref{Fig:1}. For
later use, we apply the following definitions: $\mathbf{\phi}=\left[  \phi
_{1},\phi_{2},...,\phi_{M}\right]  ^{T}$, $\mathbf{\varphi}=\left[
\varphi_{1},\varphi_{2},...,\varphi_{N}\right]  ^{T}$, $\mathbf{a}%
_{tx}=\left[  a_{tx_{1}},a_{tx_{2}},...,a_{tx_{M}}\right]  ^{T}$,
$\mathbf{a}_{rx}=\left[  a_{rx_{1}},a_{rx_{2}},...,a_{rx_{N}}\right]  ^{T},$
$\mathbf{b}_{tx}=\left[  b_{tx_{1}},b_{tx_{2}},...,b_{tx_{M}}\right]  ^{T}$
and $\mathbf{b}_{rx}=\left[  b_{rx_{1}},b_{rx_{2}},...,b_{rx_{M}}\right]
^{T}$.

An expression for the FIM $\mathbf{J}\left(  \mathbf{\psi}_{nc}\right)  ,$ is
derived in Appendix \ref{Section:appendixA}, yielding:%

\begin{equation}
\mathbf{J}{\small \left(  \mathbf{\psi}_{nc}\right)  }=\frac{2}{\sigma_{w}%
^{2}}\left[
\begin{array}
[c]{cc}%
\mathbf{S}_{nc} & \mathbf{V}_{nc}\\
\mathbf{V}_{nc}^{T} & \mathbf{\Lambda}_{\alpha}%
\end{array}
\right]  _{\left(  3MN\right)  \times\left(  3MN\right)  },\label{eq:FIMgenNC}%
\end{equation}
with the block matrices $\mathbf{S}_{nc},$ $\mathbf{\Lambda}_{\alpha},$ and
$\mathbf{V}_{nc}$ defined in the Appendix \ref{Section:appendixA} in
(\ref{e:Snc}), (\ref{e:Lambdanc}), and (\ref{e:Vnc}), respectively.

In order to determine the value of $\mathbf{J}\left(  \mathbf{\theta}%
_{nc}\right)  ,$\ we use (\ref{eq:FIMgenNC}) and (\ref{eq:Pvalue}) in
(\ref{eq:chainRule}), to obtain the following CRLB matrix:%

\begin{equation}
\mathbf{C}_{CRLB_{nc}}=\mathbf{J}^{-1}\left(  \mathbf{\theta}_{nc}\right)
=\frac{c^{2}}{2/\sigma_{w}^{2}}\left[
\begin{array}
[c]{cc}%
\mathbf{HS}_{nc}\mathbf{H}^{T} & \mathbf{HV}_{nc}\\
\mathbf{V}_{nc}^{T}\mathbf{H}^{T} & \mathbf{\Lambda}_{\alpha}%
\end{array}
\right]  ^{-1}.\label{eq:FIMlocNC}%
\end{equation}

The CRLB matrix is related to the sensor and target locations through the
matrix $\mathbf{H},$ and to the received waveforms correlation functions and
its derivatives through the $\mathbf{S}_{nc}$ and $\mathbf{V}_{nc}$ matrices.

\subsubsection{Orthogonal Waveforms}

When the waveforms are orthogonal, (\ref{e:Snc}), (\ref{e:Lambdanc}), and
(\ref{e:Vnc}) simplify to (\ref{eq:App_a9}) in Appendix
\ref{Section:appendixA}. This simplification enables to compute the CRLB
(\ref{eq:FIMlocNC}) in closed form. We perform this calculation next.

While the CRLB expresses the lower bound on the variance of the estimate of
$\mathbf{\theta}_{nc}=\left[  x,y,\mathbf{\alpha}^{R}\mathbf{,\alpha}%
^{I}\right]  ^{T}$\textbf{,} we are really interested only in the estimation
of $x $ and $y.$ The amplitude terms $\mathbf{\alpha}^{R}$ and $\mathbf{\alpha
}^{I}$ serve as nuisance parameters. For the variances of the estimates of $x$
and $y,$ it is sufficient to derive the $2\times2$ upper left submatrix
$\left[  \mathbf{C}_{CRLB_{nc}}\right]  _{2\times2}=\left[  \left(
\mathbf{J}\left(  \mathbf{\theta}_{nc}\right)  \right)  ^{-1}\right]
_{2\times2}.$

\begin{proposition}
The CRLB submatrix $\left[  \mathbf{C}_{CRLB_{nc}}\right]  _{2\times2}$ for
target localization in the \emph{non-coherent} case with orthogonal signals
is:%
\begin{equation}
\left[  \mathbf{C}_{CRLB_{nc}}\right]  _{2\times2}{\normalsize =}\frac{c^{2}%
}{2/\sigma_{w}^{2}}\left(  \mathbf{H\mathbf{S}}_{nc}\mathbf{H}^{T}\right)
^{-1}.\label{eq:CRLBexpNonCoh}%
\end{equation}

\end{proposition}

\begin{proof}
From (\ref{eq:App_a9}) in Appendix \ref{Section:appendixA}, we have for terms
of (\ref{eq:FIMlocNC}):%
\begin{align}
\mathbf{S}_{nc} &  =4\pi^{2}\beta^{2}\left[  \operatorname*{diag}%
(\mathbf{\alpha)B}\operatorname*{diag}(\mathbf{\alpha}^{\ast}\mathbf{)}%
\right]  ,\label{eq:FIMncOrto}\\
\mathbf{V}_{nc} &  =\mathbf{0},\nonumber\\
\mathbf{\Lambda}_{\alpha} &  =\mathbf{I}_{2MN\times2MN}.\nonumber
\end{align}
In (\ref{eq:FIMncOrto}), $\operatorname*{diag}(\mathbf{\alpha})$ denotes a
diagonal matrix with the elements of vector $\mathbf{\alpha}$. Matrix
$\mathbf{B}=\operatorname*{diag}\left(  \mathbf{1}\left[  \beta_{R_{1}}%
^{2},\beta_{R_{2}}^{2},...,\beta_{R_{M}}^{2}\right]  \right)  $, with
$\beta_{R_{k}}$\ denoting the normalized elements $\beta_{R_{k}}=\beta
_{k}/\beta$, and $\mathbf{1=}\left[  1,1,...1\right]  ^{T}$, $\mathbf{1}\in
R^{N\times1}$. Using (\ref{eq:FIMncOrto}) in (\ref{eq:FIMlocNC}), it is easy
to see that
\begin{align}
\left[  \mathbf{C}_{CRLB_{nc}}\right]  _{2\times2} &  =\frac{c^{2}}%
{2/\sigma_{w}^{2}}\left(  \mathbf{H\mathbf{S}}_{nc}\mathbf{H}^{T}\right)
^{-1}\label{eq:CRLBnonCohSubMatrix}\\
&  =\frac{\eta_{nc}}{g_{x_{nc}}g_{y_{nc}}-h_{nc}^{2}}\left[
\begin{array}
[c]{cc}%
g_{x_{nc}} & h_{nc}\\
h_{nc} & g_{y_{nc}}%
\end{array}
\right]  ,\nonumber
\end{align}
where:%
\begin{equation}%
\begin{array}
[c]{c}%
\eta_{nc}=\frac{c^{2}}{8\pi^{2}\beta^{2}/\sigma_{w}^{2}},\\
g_{x_{nc}}=\overset{M}{\underset{k=1}{\sum}}\overset{N}{\underset{\ell=1}%
{\sum}}\left\vert \alpha_{\ell k}\right\vert ^{2}\beta_{R_{k}}^{2}\left(
b_{tx_{k}}+b_{rx_{\ell}}\right)  ^{2},\\
g_{y_{nc}}=\overset{M}{\underset{k=1}{\sum}}\overset{N}{\underset{\ell=1}%
{\sum}}\left\vert \alpha_{\ell k}\right\vert ^{2}\beta_{R_{k}}^{2}\left(
a_{tx_{k}}+a_{rx_{\ell}}\right)  ^{2},\\
h_{nc}=-\overset{M}{\underset{k=1}{\sum}}\overset{N}{\underset{\ell=1}{\sum}%
}\left\vert \alpha_{\ell k}\right\vert ^{2}\beta_{R_{k}}^{2}\left(  a_{tx_{k}%
}+a_{rx_{\ell}}\right)  \left(  b_{tx_{k}}+b_{rx_{\ell}}\right)  .
\end{array}
\label{eq:CRLBncCoef}%
\end{equation}
This concludes the proof of the proposition.
\end{proof}

It follows that the lower bound on the variance for estimating the $x$
coordinate of the target is given by
\begin{equation}
\sigma_{x_{nc}CRB}^{2}=\eta_{nc}\frac{g_{x_{nc}}}{g_{x_{nc}}g_{y_{nc}}%
-h_{nc}^{2}}.\label{e:varx_nc}%
\end{equation}
Similarly, for the $y$ coordinate,%
\begin{equation}
\sigma_{y_{nc}CRB}^{2}=\eta_{nc}\frac{g_{y_{nc}}}{g_{x_{nc}}g_{y_{nc}}%
-h_{nc}^{2}}.\label{e:vary_nc}%
\end{equation}

The terms $g_{x_{nc}}$, $g_{y_{nc}}$, and $h_{nc}$\ are summations of
$a_{tx_{k}}$, $a_{rx_{\ell}}$, $b_{tx_{k}}$ and $b_{rx_{\ell}}$ terms that
represent sine and cosine expressions of the angles $\mathbf{\phi}$ and
$\mathbf{\varphi,}$ and therefore relate to the radars and target geometric
layout. It is apparent that for the non-coherent case, the lower bounds on the
variances (\ref{e:varx_nc}) and (\ref{e:vary_nc}) are inversely proportional
to the averaged effective bandwidth $\beta^{2}$, and $\operatorname{SNR}%
=1/\sigma_{w}^{2}$ (see expression for $\eta_{nc}$ in (\ref{eq:CRLBncCoef})).
It is interesting to note that $\eta_{nc}$ is actually the CRLB for range
estimation in a single antenna radar, based on the one-way time delay between
the radar and the target (see for example \cite{KaySSPE93}). The other terms
in (\ref{e:varx_nc}) and (\ref{e:vary_nc}) incorporate the effect of the
sensors locations.

\subsection{Coherent Processing CRLB}

\label{Section:CRLBcoherent}

We recall that in the section on the signal model, we defined the complex
amplitude $\alpha_{\ell k}$ associated with the path transmitter
$k\rightarrow$ target $\rightarrow$ receiver $\ell.$ In the non-coherent case,
the complex amplitude is a nuisance parameter in estimating the target
location $x,$ $y$. In the coherent case, the transmitting and receiving radars
are assumed to be phase synchronized. By eliminating the phase offsets, the
signal model in (\ref{e:r1}) applies, and the nuisance parameter role is left
to the complex target amplitude $\zeta=\zeta^{R}+j\zeta^{I}$. The coherent
approach to localization seeks to exploit the target location information
embedded in the phase terms $\exp\left(  -2\pi f_{c}\tau_{\ell k}\right)  $
that depend on the delays ${\tau_{\ell k},}$ which in turn are function of the
target coordinates $x,$ $y$.

Define the vector of unknown parameters:%

\begin{equation}
\mathbf{\theta}_{c}=\left[  x,y,\zeta^{R},\zeta^{I}\right]  ^{T}%
.\label{eq:theta_c}%
\end{equation}

As before, define a second vector of unknown parameters in terms of the time
delays $\mathbf{\tau}$ (rather then the target location),
\begin{equation}
\mathbf{\psi}_{c}=\left[  \mathbf{\tau},\zeta^{R},\zeta^{I}\right]
^{T},\label{eq:psi_c}%
\end{equation}
to be used in (\ref{eq:chainRule}) to derive the CRLB. In comparing the
coherent case in (\ref{eq:psi_c}) with the non-coherent counterpart in
(\ref{eq:psi_nc}), we note that $\mathbf{\psi}_{nc}$ incorporates the vectors
$\mathbf{\alpha}^{R}$ and $\mathbf{\alpha}^{I},$ while $\mathbf{\psi}_{c}$ is
a function of the scalars $\zeta^{R}$ and $\zeta^{I}.$ The reduction in the
number of unknown parameters is made possible through the measurement of the
phase terms of $\mathbf{\alpha}^{R}$ and $\mathbf{\alpha}^{I}$.

For coherent observations, the conditional, joint pdf of the observations at
the receive sensors, given by (\ref{e:r1}), is of the form:%

\begin{equation}
p\left(  \mathbf{r}|\mathbf{\psi}_{c}\right)  \varpropto\exp\left\{  -\frac
{1}{\sigma_{w}^{2}}\overset{N}{\underset{\ell=1}{\sum}}\int\limits_{T}%
\left\vert r_{\ell}(t)-\overset{M}{\underset{k=1}{\sum}}\zeta\exp\left(  -2\pi
f_{c}\tau_{\ell k}\right)  s_{k}\left(  t-\tau_{\ell k}\right)  \right\vert
^{2}dt\right\}  .\label{eq:pdf_c}%
\end{equation}

We follow the same process used in Section \ref{Section:CRLBnonCoherent}, to
develop the CRLB\ for the coherent case based on the relation in
(\ref{eq:chainRule}). The matrix $\mathbf{P}_{c}$ takes the form:%

\begin{equation}
\mathbf{P}_{c}=\frac{\partial\mathbf{\psi}_{c}}{\partial\mathbf{\theta}_{c}%
}{\small =-}\frac{1}{c}\left[
\begin{array}
[c]{cc}%
\mathbf{H} & \mathbf{0}_{MN\times2}\\
\mathbf{0}_{2\times MN} & \mathbf{I}_{2\times2}%
\end{array}
\right]  _{4\times\left(  MN+2\right)  }{\small ,}\label{eq:p_c}%
\end{equation}
where matrix $\mathbf{H}$ has the same form as in (\ref{eq:Hdef}), since it is
independent of the nuisance parameters in both cases.

An expression for the FIM matrix, $\mathbf{J}\left(  \mathbf{\psi}_{c}\right)
,$ is derived in Appendix \ref{Section:appendixB}, yielding:%

\begin{equation}
\mathbf{J}{\small \left(  \mathbf{\psi}_{c}\right)  }=\frac{2}{\sigma_{w}^{2}%
}\left[
\begin{array}
[c]{cc}%
\mathbf{S}_{c} & \mathbf{V}_{c}\\
\mathbf{V}_{c}^{T} & \mathbf{\Lambda}_{\alpha c}%
\end{array}
\right]  _{\left(  MN+2\right)  \times\left(  MN+2\right)  }%
,\label{eq:FIMgenC}%
\end{equation}
where the submatrices are found in Appendix \ref{Section:appendixB} as
follows: $\mathbf{S}_{c}$ in (\ref{e:Sc}), $\mathbf{\Lambda}_{\alpha c}$ in
(\ref{e:Lambdac}), and $\mathbf{V}_{c}$ in (\ref{e:Vc}).

The CRLB matrix for the coherent case is then found substituting
(\ref{eq:p_c})\ and (\ref{eq:FIMgenC}) in (\ref{eq:chainRule}) and
(\ref{eq:CRLB_FIM}), obtaining:%

\begin{equation}
\mathbf{C}_{CRLB_{c}}=\frac{c^{2}}{2/\sigma_{w}^{2}}\left[
\begin{array}
[c]{cc}%
\mathbf{HS}_{c}\mathbf{H}^{T} & \mathbf{HV}_{c}\\
\mathbf{V}_{c}^{T}\mathbf{H}^{T} & \mathbf{\Lambda}_{\alpha c}%
\end{array}
\right]  ^{-1}.\label{eq:FIMlocC}%
\end{equation}

As in Section \ref{Section:CRLBnonCoherent}, we develop the closed form
solution to the CRLB matrix in (\ref{eq:FIMlocC}) for the case of orthogonal
waveforms. Since we are interested only in the lower bound on the variances of
the estimates of $x$ and $y$, the submatrix $\left[  \mathbf{C}_{CRLB_{c}%
}\right]  _{2\times2}=\left[  \left(  \mathbf{J}_{c}\left(  \mathbf{\theta
}\right)  \right)  ^{-1}\right]  _{2\times2}$ is derived and evaluated next.

\begin{proposition}
The CRLB $2\times2$ submatrix for the \emph{coherent }case and orthogonal
waveforms is:%
\begin{equation}
\left[  \mathbf{C}_{CRLB_{c}}\right]  _{2\times2}{\normalsize =}\frac{c^{2}%
}{2/\sigma_{w}^{2}}\left(  \mathbf{H\mathbf{S}}_{c}\mathbf{H}^{T}%
-\mathbf{H\mathbf{V}}_{c}\mathbf{\mathbf{\Lambda}}_{\alpha c}^{-1}%
\mathbf{V}_{c}^{T}\mathbf{H}^{T}\right)  ^{-1}.\label{eq:CRLBexpCoh}%
\end{equation}

\end{proposition}

\begin{proof}
From (\ref{eq:App_b10}) in Appendix \ref{Section:appendixB} we have the values
of the matrices $\mathbf{\mathbf{S}}_{c},$ $\mathbf{\mathbf{\Lambda}}_{\alpha
c},$ and $\mathbf{\mathbf{V}}_{c}$ for orthogonal waveforms. Using this and
$\mathbf{H}$ defined in (\ref{eq:Hdef}) in (\ref{eq:FIMlocC}), the CRLB
matrix\textbf{\ \ }$\mathbf{C}_{CRLB_{c_{or}}}$ is obtained. Consequently, the
submatrix $\left[  \mathbf{C}_{CRLB_{c}}\right]  _{2\times2} $ is computed in
Appendix \ref{Section:appendixC} resulting in the form given in
(\ref{eq:CRLBexpCoh}).

This completes the proof of the proposition.
\end{proof}

From (\ref{eq:CRLBexpCoh}) and (\ref{eq:App_b10}), it can be shown that
\textbf{\ }$\left[  \mathbf{C}_{CRLB_{c}}\right]  _{2\times2}$ can be
expressed as:%

\begin{equation}
\left[  \mathbf{C}_{CRLB_{c}}\right]  _{2\times2}=\frac{\eta_{c}}{g_{x_{c}%
}g_{y_{c}}-h_{c}^{2}}\left[
\begin{array}
[c]{cc}%
g_{x_{c}} & h_{c}\\
h_{c} & g_{y_{c}}%
\end{array}
\right]  ,\label{eq:CRLBexpCohMM}%
\end{equation}
where the various quantities are as follows:%

\begin{equation}%
\begin{array}
[c]{c}%
\eta_{c}=\frac{c^{2}}{8\pi^{2}f_{c}^{2}\left(  \left\vert \zeta\right\vert
^{2}/\sigma_{w}^{2}\right)  },\\
g_{x_{c}}=\overset{M}{\underset{k=1}{\sum}}\overset{N}{\underset{\ell=1}{\sum
}}f_{R_{k}}\left(  b_{tx_{k}}+b_{rx_{\ell}}\right)  ^{2}-\frac{1}{MN}\left(
\overset{M}{\underset{k=1}{\sum}}\overset{N}{\underset{\ell=1}{\sum}}\left(
b_{tx_{k}}+b_{rx_{\ell}}\right)  \right)  ^{2},\\
g_{y_{c}}=\overset{M}{\underset{k=1}{\sum}}\overset{N}{\underset{\ell=1}{\sum
}}f_{R_{k}}\left(  a_{tx_{k}}+a_{rx_{\ell}}\right)  ^{2}-\frac{1}{MN}\left(
\overset{M}{\underset{k=1}{\sum}}\overset{N}{\underset{\ell=1}{\sum}}\left(
a_{tx_{k}}+a_{rx_{\ell}}\right)  \right)  ^{2},\\
h_{c}=-\overset{M}{\underset{k=1}{\sum}}\overset{N}{\underset{\ell=1}{\sum}%
}f_{R_{k}}\left(  a_{tx_{k}}+a_{rx_{\ell}}\right)  \left(  b_{tx_{k}%
}+b_{rx_{\ell}}\right) \\
+\frac{1}{MN}\overset{M}{\underset{k=1}{\sum}}\overset{N}{\underset{\ell
=1}{\sum}}\left(  a_{tx_{k}}+a_{rx_{\ell}}\right)  \overset{M}{\underset
{k=1}{\sum}}\overset{N}{\underset{\ell=1}{\sum}}\left(  b_{tx_{k}}%
+b_{rx_{\ell}}\right)  .
\end{array}
\label{eq:CRLBcCoef}%
\end{equation}

The lower bound on the error variance is provided by the diagonal elements of
the $\left[  \mathbf{C}_{CRLB_{c_{or}}}\right]  _{2\times2}$ submatrix and are
of the form:%

\begin{align}
\sigma_{x_{c}CRB}^{2}  &  =\eta_{c}\frac{g_{x_{c}}}{g_{x_{c}}g_{y_{c}}%
-h_{c}^{2}},\label{eq:MSEc}\\
\sigma_{y_{c}CRB}^{2}  &  =\eta_{c}\frac{g_{y_{c}}}{g_{x_{c}}g_{y_{c}}%
-h_{c}^{2}}.\nonumber
\end{align}

The terms $g_{x_{c}}$, $g_{y_{c}}$, and $h_{c}$\ are summations of $a_{tx_{k}%
}$, $a_{rx_{\ell}}$, $b_{tx_{k}}$ and $b_{rx_{\ell}}$ that represent sine and
cosine expressions of the angles $\mathbf{\phi}$ and $\mathbf{\varphi}$ and
therefore relate to the radars and target geometric layout, multiplied by the
ratio terms $f_{R_{k}}=\left(  1+\frac{\beta_{k}^{2}}{f_{c}^{2}}\right)  $.
Invoking the narrowband signals assumption $\beta_{k}^{2}/f_{c}^{2}\ll1$ it
follows that $f_{R_{k}}\simeq1$. These terms have some additional elements
when compared with the \ non-coherent case. It is apparent that for the
coherent case, the variances of the target location estimates in
(\ref{eq:MSEc})\ are inverse proportional to the carrier frequency $f_{c}^{2}$.

\subsection{Discussion}

\label{Section:DiscussionCRLB}

We make the following observations:

\begin{itemize}
\item The lower bound on the variance in the non-coherent case is inversely
proportional to the averaged effective bandwidth $\beta$. For the coherent
case, with narrowband signals, where $\beta_{k}^{2}/f_{c}^{2}\ll1$, the
localization accuracy is inversely proportional to the carrier frequency
$f_{c}$ and independent of the signal individual effective bandwidth, due to
the use of the phase information across the different paths. It is apparent
that coherent processing offers a target localization precision gain (i.e.,
reduction of the localization root mean-square error) of the order of
$f_{c}/\beta$, which we refer to as \emph{coherency gain}. Designing the ratio
$f_{c}/\beta$ to be in the range 100-1000, leads to dramatic gains.

\item The term $\eta_{c}$ in (\ref{eq:CRLBcCoef}) is the range estimate based
on one-way time delay with coherent observations for a radar with a single
antenna \cite{Minkoff02}.

\item The CRLB terms are strongly reliant on the relative geographical spread
of the radar systems vs. the target location. This dependency is incorporated
in the terms $g_{x_{nc}/x_{c}},$ $g_{y_{nc}/y_{c}}$ and $h_{nc/c}$. It is
apparent from (\ref{eq:MSEc}), (\ref{e:varx_nc}) and (\ref{e:vary_nc})\ that
there is a trade-off between the variances of the target location computed
horizontally and vertically. A set of sensor locations that minimizes the
horizontal error, may result in a high vertical error. For example, spreading
the transmitting and receiving radars in an angular range of $-(\pi/10)$ to
$+(\pi/10)$ radians with respect to the target, will result in high horizontal
error while providing low vertical error, as we would expect intuitively. This
is caused by the fact that the terms $g_{x_{nc}}/g_{x_{c}}$ are summations of
sine functions and $g_{y_{nc}}/g_{y_{c}}$ are summation of cosine functions of
the same set of angles. In order to truly determine the minimum achievable
localization accuracy in both $x$ and $y$ axis, we need to minimize the
\emph{over-all }accuracy, defined as the total variance $\sigma_{c}%
^{2}=\left(  \sigma_{x_{c}CRB}^{2}+\sigma_{y_{c}CRB}^{2}\right)  $.

\item The message of dramatic improvement in localization accuracy needs to be
moderated with the observation that the CRLB is a bound of \emph{small
errors}. As such, it ignores effects that could lead to \emph{large errors.
}For example, MIMO radar with distributed sensors and coherent observations is
subject to high sidelobes \cite{HaimBlum08}. Additionally, a phase coherent
system is sensitive to phase errors. These topics are outside the scope of
this paper, but they should be kept in perspective.

\item The lower bound as expressed by the CRLB, provides a tight bound at high
SNR, while at low SNR, the CRLB is not tight. As stated in \cite{Weiss84}, the
MLE is asymptotically unbiased and its error variance approaches the CRLB
arbitrarily close for sufficient long observation time, with the condition
that the MLE is not subject to ambiguities. As the MLE of the time estimates
is based on matched filters at the receiver end, the ambiguity features of the
signal waveforms arise in low SNR conditions and predominate the estimation
capabilities, causing erroneous time estimates. As the ambiguity problems are
usually addressed trough the signal waveform design, a more rigid bound needs
to be found for the localization variance in the low SNR case.
\end{itemize}

\section{Effect of Sensors Locations}

\label{Section:OptimizationOverAll}

The CRLB\ for target localization with coherent MIMO\ radar shows a gain,
i.e., reduction in the standard deviation of the localization estimate, of
$f_{c}/\beta$ compared to non-coherent localization. Yet, the CRLB is strongly
dependent on the locations of the transmitting and receiving sensors relative
to the target location, through the terms $g_{x_{nc}/x_{c}}, $ $g_{y_{nc}%
/y_{c}}$ and $h_{nc/c}$. To gain a better understanding of these relations,
and set a lower bound on the CRLB over all possible sensor placements, further
analysis is developed in this section.

We introduce the following general notation: for any given set of vectors
$\mathbf{\xi}=\left(  \xi_{1},\xi_{2},...,\xi_{L}\right)  $\ and
$\mathbf{\kappa}=\left(  \kappa_{1},\kappa_{2},...,\kappa_{L}\right)  $:%

\begin{equation}%
\begin{array}
[c]{c}%
T\left(  \mathbf{\xi}\right)  =\frac{1}{L}\overset{L}{\underset{i=1}{\sum}}%
\xi_{i}\\
T\left(  \mathbf{\xi}^{2}\right)  =\frac{1}{L}\overset{L}{\underset{i=1}{\sum
}}\xi_{i}^{2}\\
T\left(  \mathbf{\xi\kappa}\right)  =\frac{1}{L}\overset{L}{\underset
{i=1}{\sum}}\xi_{i}\kappa_{i}.
\end{array}
\label{eq:GeneralOperators}%
\end{equation}

The terms $g_{x_{c}}$ and $g_{y_{c}}$ in (\ref{eq:CRLBncCoef})\ can be
expressed using the conventions defined in (\ref{eq:GeneralOperators}) and
terms defined in Section \ref{Section:CRLBcoherent}, viz.:
\begin{equation}
g_{x_{c}}=MN\left[  T\left(  \mathbf{b}_{tx}^{2}\right)  +T\left(
\mathbf{b}_{rx}^{2}\right)  -\left[  T\left(  \mathbf{b}_{tx}\right)  \right]
^{2}-\left[  T\left(  \mathbf{b}_{rx}\right)  \right]  ^{2}\right]
,\label{eq:gxCoh2}%
\end{equation}
and
\begin{equation}
g_{y_{c}}=MN\left[  T\left(  \mathbf{a}_{tx}^{2}\right)  +T\left(
\mathbf{a}_{rx}^{2}\right)  -\left[  T\left(  \mathbf{a}_{tx}\right)  \right]
^{2}-\left[  T\left(  \mathbf{a}_{rx}\right)  \right]  ^{2}\right]
,\label{eq:gyCoh2}%
\end{equation}
where the narrowband signals assumption is applied. Similarly, the term
$h_{c}$ in (\ref{eq:CRLBcCoef}) can be expressed:%

\begin{align}
h_{c}  &  =MN\left[  T\left(  \mathbf{a}_{tx}\mathbf{b}_{tx}\right)  +T\left(
\mathbf{a}_{rx}\mathbf{b}_{rx}\right)  \right. \label{eq:hCoh2}\\
&  \left.  -T\left(  \mathbf{a}_{tx}\right)  E\left(  \mathbf{b}_{tx}\right)
-T\left(  \mathbf{a}_{rx}\right)  E\left(  \mathbf{b}_{rx}\right)  \right]
.\nonumber
\end{align}
Since $a_{tx_{k}}^{2}+b_{tx_{k}}^{2}=\cos^{2}\phi_{_{k}}+\sin^{2}\phi_{_{k}%
}=1$ and $a_{rx_{\ell}}^{2}+b_{rx_{\ell}}^{2}=\cos^{2}\varphi_{\ell}+\sin
^{2}\varphi_{\ell}=1,$ the following conditions apply:
\begin{equation}%
\begin{array}
[c]{c}%
T\left(  \mathbf{a}_{tx}^{2}\right)  +T\left(  \mathbf{b}_{tx}^{2}\right)
=1\\
T\left(  \mathbf{a}_{rx}^{2}\right)  +T\left(  \mathbf{b}_{rx}^{2}\right)
=1\\
0\leq\left[  T\left(  \mathbf{a}_{tx}\right)  \right]  ^{2}\leq1;\text{ }%
0\leq\left[  T\left(  \mathbf{a}_{rx}\right)  \right]  ^{2}\leq1\\
0\leq\left[  T\left(  \mathbf{b}_{tx}\right)  \right]  ^{2}\leq1;\text{ }%
0\leq\left[  T\left(  \mathbf{b}_{rx}\right)  \right]  ^{2}\leq1\\
0\leq T\left(  \mathbf{a}_{tx}^{2}\right)  \leq1;\text{ }0\leq T\left(
\mathbf{a}_{rx}^{2}\right)  \leq1\\
0\leq T\left(  \mathbf{b}_{tx}^{2}\right)  \leq1;\text{ }0\leq T\left(
\mathbf{b}_{rx}^{2}\right)  \leq1.
\end{array}
\label{eq:rel2}%
\end{equation}

We seek to find sets of angles $\mathbf{\phi}^{\ast}$ and $\mathbf{\varphi
}^{\mathbf{\ast}},$ that yield sets of cosine and sine expressions
$\mathbf{a}_{tx}^{\ast},\mathbf{a}_{rx}^{\ast},\mathbf{b}_{tx}^{\ast
},\mathbf{b}_{rx}^{\ast}$ for which the values of the Cramer-Rao bounds for
localization along the $x$ and $y$ axes ($\sigma_{x_{c}CRB}^{2}$ and
$\sigma_{y_{c}CRB}^{2},$ respectively) are jointly minimized, that is:%

\begin{equation}%
\begin{array}
[c]{c}%
\underset{\mathbf{a}_{tx},\mathbf{a}_{rx},\mathbf{b}_{tx},\mathbf{b}_{rx}%
}{\text{minimize}}\left(  \sigma_{x_{c}CRB}^{2}+\sigma_{y_{c}CRB}^{2}\right)
.
\end{array}
\label{eq:min1}%
\end{equation}
This is equivalent to minimizing the trace of the CRLB submatrix $\left[
\mathbf{C}_{CRLB_{c}}\right]  _{2\times2}$. The explicit minimization problem
is formulated introducing the objective function $f_{0}$:\bigskip%
\begin{equation}%
\begin{array}
[c]{cc}%
\underset{\mathbf{a}_{tx},\mathbf{a}_{rx},\mathbf{b}_{tx},\mathbf{b}_{rx}%
}{\text{minimize}} & f_{0}\left(  \mathbf{a}_{tx},\mathbf{a}_{rx}%
,\mathbf{b}_{tx},\mathbf{b}_{rx}\right)  =\eta_{c}\frac{g_{x_{c}}+g_{y_{c}}%
}{g_{x_{c}}g_{y_{c}}-h_{c}^{2}}\\
\text{\ } & \text{subject to constraints (\ref{eq:rel2}).}%
\end{array}
\label{eq:min2}%
\end{equation}

This representation of the problem is not a convex optimization
problem.\footnote{A convex optimization problem is of the form \cite{Boyd}
\par%
\[%
\begin{array}
[c]{cc}%
\text{minimize} & f_{0}\left(  x\right) \\
\text{subject to} & f_{i}\left(  x\right)  \leq0\\
& \sum_{j}a_{j}x_{j}=0
\end{array}
\]
for some constants $a_{i},$ $i,$ $j,$ $i=1,...,m,$ $j$ $=1,...,p,$ and where
$f_{0},...,f_{m}$ are convex functions.} The next steps are undertaken in
order to formulate a convex optimization problem equivalent to (\ref{eq:min2}%
), i.e., a convex optimization problem that can be solved through routine
techniques and from whose solution it is readily possible to find the solution
to (\ref{eq:min2}).

In \cite{Lee75}, it is shown that for a given positive definite matrix, in our
case $\left[  \mathbf{C}_{CRLB_{c}}\right]  _{2\times2}$, and its inverse
matrix $\mathbf{F}$, in this case:%

\begin{equation}
\mathbf{F}=\frac{1}{\eta_{c}}\left[
\begin{array}
[c]{cc}%
g_{yc} & -h_{c}\\
-h_{c} & g_{xc}%
\end{array}
\right]  ,\label{eq:F}%
\end{equation}
the following relation exists between the diagonal elements of these
matrices:
\begin{equation}
\left[  \mathbf{C}_{CRLB_{c}}\right]  _{ii}\geq\frac{1}{\left[  \mathbf{F}%
\right]  _{ii}};\text{ }i=1,2.\label{eq:ineq1}%
\end{equation}
Equality conditions apply for all $i$ iff $\mathbf{F}$ is a diagonal matrix,
i.e., $h_{c}=0$. Now, observe that the inverse of the elements on the diagonal
of $\mathbf{F}$\ are lower bounding the elements on the diagonal of the matrix
$\mathbf{C}_{CRLB_{c}}$\ for any $\mathbf{a}_{tx},\mathbf{a}_{rx}%
,\mathbf{b}_{tx},\mathbf{b}_{rx}$\textbf{.} We then define the objective
function $\overline{f_{0}}\left(  \mathbf{a}_{tx},\mathbf{a}_{rx}%
,\mathbf{b}_{tx},\mathbf{b}_{rx}\right)  ,$ and the optimization problem
\begin{align}
\text{min }\overline{f_{0}}\left(  \mathbf{a}_{tx},\mathbf{a}_{rx}%
,\mathbf{b}_{tx},\mathbf{b}_{rx}\right)   &  =\frac{1}{\eta_{c}}\left(
\frac{1}{g_{x_{c}}}+\frac{1}{g_{y_{c}}}\right) \label{eq:min21}\\
&  \text{subject to (\ref{eq:rel2}).}\nonumber
\end{align}
The new objective function and the original objective function are related as
$f_{0}\left(  \mathbf{a}_{tx},\mathbf{a}_{rx},\mathbf{b}_{tx},\mathbf{b}%
_{rx}\right)  \geq$ $\overline{f_{0}}\left(  \mathbf{a}_{tx},\mathbf{a}%
_{rx},\mathbf{b}_{tx},\mathbf{b}_{rx}\right)  $, with equality for $h_{c}=0$.
Substitute the values of $g_{x_{c}}$ and $g_{y_{c}}$ from (\ref{eq:gxCoh2})
and (\ref{eq:gyCoh2}) in the objective function of (\ref{eq:min21}) to obtain%

\begin{align}
\overline{f_{0}}\left(  \mathbf{a}_{tx},\mathbf{a}_{rx},\mathbf{b}%
_{tx},\mathbf{b}_{rx}\right)   &  =\frac{1/\left(  \eta_{c}MN\right)
}{2-T\left(  \mathbf{b}_{tx}^{2}\right)  -T\left(  \mathbf{b}_{rx}^{2}\right)
-\left[  T\left(  \mathbf{a}_{tx}\right)  \right]  ^{2}-\left[  T\left(
\mathbf{a}_{rx}\right)  \right]  ^{2}}\label{eq:obj1}\\
&  +\frac{1/\left(  \eta_{c}MN\right)  }{T\left(  \mathbf{b}_{tx}^{2}\right)
+T\left(  \mathbf{b}_{rx}^{2}\right)  -\left[  T\left(  \mathbf{b}%
_{tx}\right)  \right]  ^{2}-\left[  T\left(  \mathbf{b}_{rx}\right)  \right]
^{2}}.\nonumber
\end{align}
\qquad It is apparent that the denominator of the first summand is bounded by:%

\begin{equation}
0\leq2-T\left(  \mathbf{b}_{tx}^{2}\right)  -T\left(  \mathbf{b}_{rx}%
^{2}\right)  -\left[  T\left(  \mathbf{a}_{tx}\right)  \right]  ^{2}-\left[
T\left(  \mathbf{a}_{rx}\right)  \right]  ^{2}\leq2-T\left(  \mathbf{b}%
_{tx}^{2}\right)  -T\left(  \mathbf{b}_{rx}^{2}\right)  ,\label{eq:ineq22}%
\end{equation}
and the denominator of the second summand is bounded by:%

\begin{equation}
0\leq T\left(  \mathbf{b}_{tx}^{2}\right)  +T\left(  \mathbf{b}_{rx}%
^{2}\right)  -\left[  T\left(  \mathbf{b}_{tx}\right)  \right]  ^{2}-\left[
T\left(  \mathbf{b}_{rx}\right)  \right]  ^{2}\leq T\left(  \mathbf{b}%
_{tx}^{2}\right)  +T\left(  \mathbf{b}_{rx}^{2}\right)  .\label{eq:ineq23}%
\end{equation}
Denote $T\left(  \mathbf{b}_{tx}^{2}\right)  +T\left(  \mathbf{b}_{rx}%
^{2}\right)  =\mu$, and let $T\left(  \mathbf{a}_{tx}\right)  =T\left(
\mathbf{a}_{rx}\right)  =T\left(  \mathbf{b}_{tx}\right)  =T\left(
\mathbf{b}_{rx}\right)  =0.$ Then, from (\ref{eq:obj1})-(\ref{eq:ineq23}) and
(\ref{eq:min21}), we obtain the following problem:%
\begin{equation}%
\begin{array}
[c]{cc}%
\underset{\mathbf{\mu}}{\text{minimize}} & \overline{f_{0}}\left(  \mu\right)
=\dfrac{1}{2-\mu}+\dfrac{1}{\mu}\\
\text{subject to} & \mu-2\leq0\\
& -\mu\leq0.
\end{array}
\label{eq:obj21}%
\end{equation}

The objective function $\overline{f_{0}}\left(  \mu\right)  $ is still not
convex. The epigraph form is a way to introduce a linear (and hence convex)
objective $t$, while the original objective $\overline{f_{0}}\ $is
incorporated into a new constraint $\overline{f_{0}}-t\leq0.$ The key point
here is that while $\overline{f_{0}}$ is not convex, the constraint$\overline
{f_{0}}-t\leq0$ can be transformed to a convex form. After some simple
algebraic manipulations, the epigraph form turns into the following convex problem:%

\begin{equation}%
\begin{array}
[c]{cc}%
\underset{\mu,t}{\text{minimize}} & t\\
\text{subject to\ } & \\
&
\begin{array}
[c]{c}%
t\mu^{2}-2t\mu+2\leq0\\
\mu-2\leq0\\
-\mu\leq0\\
-t\leq0
\end{array}
.
\end{array}
\label{eq:min4}%
\end{equation}
\qquad

A convenient way to solve this convex optimization problem is to employ the
concept of Lagrange duality and exploit the sufficiency of the
\textit{Karusk-Kuhn-Tucker }(KKT)\ conditions \cite{Boyd}. The Lagrangian of
the problem in (\ref{eq:min4}) is given by:%

\begin{equation}
L(\mu,t,\mathbf{\lambda})=t+\lambda_{1}\left(  t\mu^{2}-2t\mu+2\right)
+\lambda_{2}\left(  \mu-2\right)  -\lambda_{3}\mu-\lambda_{4}%
t,\label{eq:Lagra}%
\end{equation}
where $\lambda_{i}$, $i=1,..,4$ is the\textit{\ Lagrange multiplier}
associated with the $i$th inequality constraint $f_{i}\left(  \mu,t\right)
\leq0$.

The KKT conditions state that the optimal solution for the primal problem
(minimization of $t$ in (\ref{eq:min4})) is given by the solution to the set
of equations:%
\begin{align}
\frac{\partial L(\mu,t,\mathbf{\lambda})}{\partial\mu} &  =0\label{eq:KKT}\\
\frac{\partial L(\mu,t,\mathbf{\lambda})}{\partial t} &  =0\nonumber\\
f_{i}(\mu,t) &  \leq0;\text{ \ }i=1,..,4\nonumber\\
\lambda_{i} &  \geq0;\text{ \ }i=1,..,4\nonumber\\
\lambda_{i}f_{i}(\mu,t) &  =0;\text{ \ }i=1,..,4.\nonumber
\end{align}
Applied to (\ref{eq:min4}) and (\ref{eq:Lagra}), these equations specialize to%
\begin{align}
\lambda_{1}\left(  2t\mu-2t\right)  +\lambda_{2}-\lambda_{3} &
=0\label{eq:KKT1}\\
1+\lambda_{1}\left(  \mu^{2}-2\mu\right)  -\lambda_{4} &  =0\nonumber\\
\lambda_{1}\left(  t\mu^{2}-2t\mu+2\right)   &  =0\nonumber\\
\lambda_{2}\left(  \mu-2\right)   &  =0\nonumber\\
-\lambda_{3}\mu &  =0\nonumber\\
-\lambda_{4}t &  =0.\nonumber
\end{align}
It is not difficult to show that the solution to this system is given by
\begin{equation}%
\begin{array}
[c]{c}%
\mu^{\ast}=1\\
t^{\ast}=2\\
\lambda_{1}^{\ast}=1\\
\lambda_{2}^{\ast}=\lambda_{3}^{\ast}=\lambda_{4}^{\ast}=0
\end{array}
.\label{eq:opt}%
\end{equation}

Recalling that $\mu=T\left(  \mathbf{b}_{tx}^{2}\right)  +T\left(
\mathbf{b}_{rx}^{2}\right)  ,$ the optimal solution can be rewritten as:%

\begin{equation}
\mu^{\ast}=T\left(  \mathbf{b}_{tx}^{\ast2}\right)  +T\left(  \mathbf{b}%
_{rx}^{\ast2}\right)  =1.\label{eq:OptimalSet2}%
\end{equation}
In addition to (\ref{eq:OptimalSet2}), $\mathbf{a}_{tx}^{\ast},\mathbf{a}%
_{rx}^{\ast},\mathbf{b}_{tx}^{\ast},\mathbf{b}_{rx}^{\ast}$ have to satisfy
the relations (\ref{eq:rel2}), and the equality conditions for (\ref{eq:ineq1}%
), (\ref{eq:ineq22}) and (\ref{eq:ineq23}), viz.,%

\begin{equation}%
\begin{array}
[c]{c}%
T\left(  \mathbf{a}_{tx}^{\ast2}\right)  +T\left(  \mathbf{a}_{rx}^{\ast
2}\right)  =1\\
T\left(  \mathbf{b}_{tx}^{\ast}\right)  =0;\text{ }T\left(  \mathbf{b}%
_{rx}^{\ast}\right)  =0\\
T\left(  \mathbf{a}_{tx}^{\ast}\right)  =0;\text{ }T\left(  \mathbf{a}%
_{rx}^{\ast}\right)  =0\\
T\left(  \mathbf{a}_{tx}^{\ast}\mathbf{b}_{tx}^{\ast}\right)  +T\left(
\mathbf{a}_{rx}^{\ast}\mathbf{b}_{rx}^{\ast}\right)  =0.
\end{array}
\label{eq:OptimalSet3}%
\end{equation}
Substituting these results in (\ref{eq:gxCoh2}) and (\ref{eq:gyCoh2}), we
compute the optimal $g_{x_{c}}^{\ast}$ and $g_{y_{c}}^{\ast},$
\[
g_{x_{c}}^{\ast}=g_{y_{c}}^{\ast}=MN.
\]
It follows that the minimum value of the trace of the Cramer Rao matrix
$\left[  \mathbf{C}_{CRLB_{c_{or}}}\right]  _{_{2\times2}},$ $f_{0}$ in
(\ref{eq:min2}), is given by:%

\begin{equation}
f_{0}\left(  \mathbf{a}_{tx}^{\ast},\mathbf{a}_{rx}^{\ast},\mathbf{b}%
_{tx}^{\ast},\mathbf{b}_{rx}^{\ast}\right)  =\frac{2\eta_{c}}{MN}%
.\label{eq:opttrace}%
\end{equation}

The final step in determining the effect of sensor locations on the
localization CRLB is to recall that the multivariable argument of $f_{0}$ in
(\ref{eq:opttrace}) is actually a function of the transmitting sensors angles
$\phi_{k},$ $k=1,\ldots,M,$ and receiving sensors angles $\varphi_{\ell},$
$\ell=1,\ldots,N$ (see definitions in the previous section). What are then the
optimal sets $\mathbf{\phi}^{\ast}$ and $\mathbf{\varphi}^{\ast}$ that
minimize the variance of the localization error?\ The optimal angles can be
found from the relations (\ref{eq:OptimalSet3}). For example, for the cosine
of the transmitters bearings $T\left(  \mathbf{a}_{tx}^{\ast}\right)  =0,$
means
\begin{equation}
\frac{1}{M}\overset{M}{\underset{k=1}{\sum}}\cos\phi_{k}^{\ast}%
=0.\label{e:sum1}%
\end{equation}
A symmetrical set of angles of the form $\mathbf{\phi}^{\ast}=\left\{
\phi_{i}^{\ast}|\phi_{i}^{\ast}=\phi_{_{0}}+\frac{2\pi\left(  i-1\right)  }%
{M};i=1,..,M;M\geq2\right\}  $, is a solution to (\ref{e:sum1}) for any
arbitrary $\phi_{_{0}}.$ The same solution is obtained for the sines,
$T\left(  \mathbf{b}_{tx}^{\ast}\right)  =0.$ The relations $T\left(
\mathbf{a}_{rx}^{\ast}\right)  =0,$ $T\left(  \mathbf{b}_{rx}^{\ast}\right)
=0$ lead to a solution constituted by a symmetrical set of angles
$\mathbf{\varphi}^{\ast}$ of the same form as $\mathbf{\phi}^{\ast}.$ The
relation $T\left(  \mathbf{a}_{tx}^{\ast}\mathbf{b}_{tx}^{\ast}\right)
+T\left(  \mathbf{a}_{rx}^{\ast}\mathbf{b}_{rx}^{\ast}\right)  =0$ expressed
in terms of angles is
\begin{equation}
\frac{1}{M}\sum_{k=1}^{M}\cos\phi_{k}^{\ast}\sin\phi_{k}^{\ast}+\frac{1}%
{N}\sum_{\ell=1}^{N}\cos\varphi_{\ell}^{\ast}\sin\varphi_{\ell}^{\ast
}=0.\label{e:sum2}%
\end{equation}
It can be shown that (\ref{e:sum2}) is met by angles $\phi_{k}^{\ast}$ and
$\varphi_{\ell}^{\ast}$ symmetrically distributed around the unit circle, but
the number of sensors has to meet $M\geq3,$ $N\geq3.$ The condition $T\left(
\mathbf{b}_{tx}^{\ast2}\right)  +T\left(  \mathbf{b}_{rx}^{\ast2}\right)  =1$
in (\ref{eq:OptimalSet3}), expressed in its explicit form, is%

\begin{equation}
\frac{1}{M}\sum_{k=1}^{M}\cos^{2}\phi_{k}^{\ast}+\frac{1}{N}\sum_{\ell=1}%
^{N}\cos^{2}\varphi_{\ell}^{\ast}=1.\label{e:sum3}%
\end{equation}
The symmetrical set of angles that meet (\ref{e:sum1}) and (\ref{e:sum2})
provide $\frac{1}{M}\sum_{k=1}^{M}\cos^{2}\phi_{k}^{\ast}=$ $\frac{1}{N}%
\sum_{\ell=1}^{N}\cos^{2}\varphi_{\ell}^{\ast}=\frac{1}{2}$ and therefore meet
the requirement of (\ref{e:sum3}). The same applies to $T\left(
\mathbf{a}_{tx}^{\ast2}\right)  +T\left(  \mathbf{a}_{rx}^{\ast2}\right)  =1$
, where we have $\frac{1}{M}\sum_{k=1}^{M}\sin^{2}\phi_{k}^{\ast}=$ $\frac
{1}{N}\sum_{\ell=1}^{N}\sin^{2}\varphi_{\ell}^{\ast}=\frac{1}{2}$.

We conclude that $M\geq3$ transmitting, and $N\geq3$ receiving sensors,
symmetrically placed on a circle around the target at angular spacings of
$2\pi/M$ and $2\pi/N,$ respectively, lead to the lowest value of the
localization CRLB.

This result can be extended by noticing that relations (\ref{eq:OptimalSet3})
also hold for any \emph{superposition} of symmetrical sets containing no less
than $3$ transmitting and/or receiving sensors. Therefore, the complete set of
optimal points is given by:%

\begin{equation}%
\begin{array}
[c]{c}%
\mathbf{\phi}^{\ast}=\left\{  \phi_{k}^{\ast}\left\vert \left.  \left(
\phi_{k}^{\ast}=\phi_{_{v}}+\frac{2\pi\left(  z-1\right)  }{Z_{v}}\right)
\right\vert _{z=1,..,Z_{v}}\right.  ;Z_{v}\geq3;\underset{v=1}{\overset{V}{%
{\displaystyle\sum}
}}Z_{v}=M\text{ }\ \right\} \\
\mathbf{\varphi}^{\ast}=\left\{  \varphi_{\ell}^{\ast}\left\vert \left.
\left(  \varphi_{\ell}^{\ast}=\varphi_{_{u}}+\frac{2\pi\left(  z-1\right)
}{Z_{u}}\right)  \right\vert _{z=1,..,Z_{u}}\right.  ;Z_{u}\geq3;\underset
{u=1}{\overset{U}{%
{\displaystyle\sum}
}}Z_{u}=N\text{ }\right\}  ,
\end{array}
\label{e:optSet}%
\end{equation}
where the total number of transmitting ($M$) and receiving ($N$) radars may be
divided into $V$ and $U$\ sets of symmetrically placed radars, each set
consists of $Z_{v}$ and $Z_{u}$ radars, respectively. The angles $\phi_{_{v}}$
and $\varphi_{_{u}}$ are an initial arbitrary rotation of the symmetric sets
$Z_{v}$ and $Z_{u}$, correspondingly.

As a special case, it is interesting to evaluate the CRLB in
(\ref{eq:CRLBexpCohMM}) with $1$ transmitter and $MN$\ receivers, i.e., a
Single-Input Multiple-Output (SIMO) system. This scheme makes use of $\left(
MN+1\right)  $ radars instead of $\left(  M+N\right)  $ radars used in a MIMO
system with $M$ transmitters and $N$ receivers. From (\ref{e:optSet}) it is
apparent the this case does not provide optimality since the number of
transmitters is smaller than $3$. To evaluate $\sigma_{x_{c}CRB}^{2}%
+\sigma_{y_{c}CRB}^{2}$ for this setting we assume $1$ transmitter is located
at an arbitrary angle $\phi_{_{1}}$ with respect to the target, and a set of
$MN$ receivers are located symmetrically around the target, at angles
$\mathbf{\varphi}^{\ast}$ that follow the condition in (\ref{e:optSet}). The
expressions in (\ref{eq:gxCoh2}), (\ref{eq:gyCoh2}), and (\ref{eq:hCoh2})
reduce to the form:
\begin{align}
g_{x_{c}} &  =MN\left[  T\left(  \mathbf{b}_{rx}^{2}\right)  -\left[  T\left(
\mathbf{b}_{rx}\right)  \right]  ^{2}\right]  =\frac{1}{2}%
MN,\label{eq:SIMOcoef}\\
g_{y_{c}} &  =MN\left[  T\left(  \mathbf{a}_{rx}^{2}\right)  -\left[  T\left(
\mathbf{a}_{rx}\right)  \right]  ^{2}\right]  =\frac{1}{2}MN,\nonumber\\
h_{c} &  =MN\left[  T\left(  \mathbf{a}_{rx}\mathbf{b}_{rx}\right)  -T\left(
\mathbf{a}_{rx}\right)  E\left(  \mathbf{b}_{rx}\right)  \right]  =0,\nonumber
\end{align}
and the trace of the CRLB submatrix $\left[  \mathbf{C}_{CRLB_{c}}\right]
_{2\times2}$, defined by $f_{0}\left(  \mathbf{a}_{tx},\mathbf{a}%
_{rx},\mathbf{b}_{tx},\mathbf{b}_{rx}\right)  =\sigma_{x_{c}CRB}^{2}%
+\sigma_{y_{c}CRB}^{2}=\eta_{c}\frac{g_{x_{c}}+g_{y_{c}}}{g_{x_{c}}g_{y_{c}%
}-h_{c}^{2}}$, is
\begin{equation}
f_{0}\left(  \mathbf{a}_{tx},\mathbf{a}_{rx},\mathbf{b}_{tx},\mathbf{b}%
_{rx}\right)  =\frac{4\eta_{c}}{MN}.\label{eq:SIMOtrace}%
\end{equation}
This result expresses an increase in the estimation error in the factor
of\ $2$ when compared with $M$ transmitters and $N$ receivers given in
(\ref{eq:opttrace}).

\subsection{Discussion}

\label{Section:DiscussionOpt}

The following comments are intended to provide further insight into the
results obtained in this section.

\begin{itemize}
\item From (\ref{eq:opttrace}), the lowest CRLB for target localization
utilizing phase information is given by $2\eta_{c}/\left(  MN\right)  $. We
interpret the reduction of the CRLB by the factor $MN/2$ compared to a single
antenna range estimation given by $\eta_{c}$ as a \emph{MIMO radar gain.} This
gain reflects two effects: (1) the gain due to the system footprint; (2) the
advantage of using $M$ transmitters and $N$ receivers, rather than, for
example, $1$ transmitter and $MN$ receivers. The latter gain is apparent when
$MN\gg\left(  M+N\right)  $.

\item The CRLB obtained through the use of a single transmit antenna and $MN$
receive antennas in (\ref{eq:SIMOtrace}) is $4\eta_{c}/\left(  MN\right)  $.
It follows that MIMO\ radar, with a total of $M+N$ sensors, has twice the
performance (from the point of view of localization CRLB) of a system with a
single transmit antenna and $MN$ receive antennas.

\item The best accuracy is obtained when the transmitting and receiving radars
are located on a virtual circle, centered at the target position, with uniform
angular spacings of $2\pi/M$ and $2\pi/N$, respectively, or any
\emph{superposition} of such sets.

\item The optimization analysis presented in this section is intended to
provide insight into the effect the sensors locations have on the CRLB.
Naturally, in practice, it is not possible to control in real time the
location of the sensors relative to a target. However, the results here teach
us that selecting among the sensors those who are most symmetrical with
respect to the target may lead to the most accurate localization.
\end{itemize}

So far we have focused on the theoretical lower bound of the localization
error. In the next section, we discuss specific techniques for target
localization and their performance as a function of sensors locations. For
this purpose, the GDOP metric and GDOP\ contour mapping tools are introduced.

\section{Methods for Target Localization}

\label{Section:TargetEst}

In Section \ref{Section:CRLB}, was formulated the lower bound on the variance
of any localization estimate. Here, it is of interest to discuss some specific
target localization estimators. In particular, two estimators are presented:
the MLE and the BLUE. The MLE is motivated by its asymptotic optimality, while
the BLUE by its closed form expression.

\subsection{MLE Target Localization}

\label{Section:TargetEst MLE}

The MLE is a practical estimator in the sense that its application to a
problem of observations in white Gaussian noise is relatively straightforward.
Moreover, under mild conditions on the probability density function of the
observations, the MLE of the unknown parameters is asymptotically unbiased,
and it asymptotically attains the CRLB \cite{KaySSPE93}.

For the case of coherent MIMO radar, the signal waveform received by radar
$\ell$ is given in (\ref{e:r}). The MLE of the unknown parameter vector
$\mathbf{\theta}=\left[  x,y,\zeta\right]  ^{T}$ given the observation vector
$\mathbf{r}$ is given by \cite{KaySSPE93}:%

\begin{equation}
\widehat{\mathbf{\theta}}_{_{ML}}=\arg\left\{  \underset{\mathbf{\theta}}%
{\max}\left[  \log p\left(  \mathbf{r}|\mathbf{\theta}\right)  \right]
\right\}  ,\label{eq:12}%
\end{equation}
where $p\left(  \mathbf{r}|\mathbf{\theta}\right)  $ is given by
(\ref{eq:pdf_c}) noting that the time delays $\tau_{\ell k}$ are known
functions of $x$ and $y$. To jointly maximize $\log p\left(  \mathbf{r}%
|\mathbf{\theta}\right)  $ with respect to $\mathbf{\theta}=\left[
x,y,\zeta\right]  ^{T},$ we start by maximizing it with respect to $\zeta$:%

\begin{equation}
\frac{\partial}{\partial\zeta}\log p\left(  \mathbf{r}|x,y,\zeta\right)
\mid_{\zeta=\widehat{\zeta}}=0.\label{eq:13}%
\end{equation}
Using (\ref{eq:pdf_c}) in (\ref{eq:13}), the estimate $\widehat{\zeta}$ can be
found, and it is a function of $x$ and $y.$ By substituting it back into
(\ref{eq:12}), it is said to \emph{compress }the log-likelihood function
\cite{Haykin04} to $\log p\left(  \mathbf{r}|x,y,\widehat{\zeta}\right)  $.
The MLE\ of the target location is then given by
\begin{align}
\frac{\partial}{\partial x}\log p\left(  \mathbf{r}|x,y,\widehat{\zeta
}\right)   &  \mid_{x=\widehat{x}_{ML}}=0\nonumber\\
\frac{\partial}{\partial y}\log p\left(  \mathbf{r}|x,y,\widehat{\zeta
}\right)   &  \mid_{y=\widehat{y}_{ML}}=0.\label{e:MLE}%
\end{align}
Since a closed form expression can not be found for the MLE in (\ref{e:MLE}),
numerical methods need to be applied. A grid search or an iterative
maximization of the likelihood function needs to be performed to determine
$\widehat{x}_{ML}$ and $\widehat{y}_{ML}$. This might involve a significant
computational effort. In practice, we can limit the search grid for high
resolution target localization estimation to an area around a coarse initial
estimate obtained by the non-coherent approach.

\subsection{ BLUE Target Localization}

\label{Section:TargetEst BLUE}

The MLE\ presented in Section (\ref{Section:TargetEst MLE}) does not lend
itself to a closed form expression, and numerical methods need to be used to
solve it. A closed form solution to the target localization can be obtained by
application of the BLUE.

To formulate the BLUE, it is necessary to have an observation model in which
observations change linearly with the target location coordinates. That is
because it is inherent to the BLUE that the estimate is \emph{linear}. To this
end, we formulate a model in which the time delays are \textquotedblleft
observable.\textquotedblright\ Let the observed time delay associated with a
transmitter-receiver pair be $\mu_{\ell k},$ then
\begin{equation}
\mu_{\ell k}=\tau_{\ell k}+\varepsilon_{\ell k},\ \ \ \ \ \forall
k=1,..,M,l=1,..,N,\label{e:mu}%
\end{equation}
where $\varepsilon_{\ell k}$ is the \textquotedblleft observation
noise.\textquotedblright\ In practice, the time delays are not directly
observable. Rather, they are estimated, for example by maximum likelihood,
from the received signals. Then, the term $\varepsilon_{\ell k}$ is the time
delay estimation error. Our BLUE estimation problem of the target location
should not be confused with the estimation of the time delays. The estimation
of the time delays is just a preparatory step in setting up the
\textquotedblleft observations\textquotedblright\ of the BLUE model. Once, the
observation model has been set up, it is necessary to ensure that the model
between the time delays and target location is linear. Setting the origin of
the coordinate system at some nominal estimate of the target location, and
preserving only linear terms of the Taylor expansion of expressions such as in
(\ref{e:tau_vq}), we can express the time delays\ as linear functions of $x$
and $y,$
\begin{equation}
\tau_{\ell k}\approx-\frac{x}{c}\left(  \cos\phi_{k}+\cos\varphi_{\ell
}\right)  -\frac{y}{c}\left(  \sin\phi_{k}+\sin\varphi_{\ell}\right)
,\label{eq:16}%
\end{equation}
where the angles $\phi_{k}$ and $\varphi_{\ell}$\ are the bearings that the
transmitting sensor $k$ and receiving sensor $\ell,$ respectively, subtend
with the reference axis (with the origin at the nominal estimate of the target
location). Note that the definitions of the angles here are a little different
than the angles defined in Section \ref{Section:CRLB} and also denoted $\phi$
and $\varphi.$ Here, the vertex of the angles is an arbitrary point in the
neighborhood of the true target location. In Section \ref{Section:CRLB}, the
vertex is at the true target location. Since only the vertex is different, we
preserved the same notation for simplicity sake. Utilizing definitions
(\ref{eq:abDef}), we can express the linear model in the following simplified form:%

\begin{equation}
\tau_{\ell k}=-\frac{x}{c}\left(  a_{tx_{k}}+a_{rx_{\ell}}\right)  -\frac
{y}{c}\left(  b_{tx_{k}}+b_{rx_{\ell}}\right)  .\label{eq:17}%
\end{equation}

Letting, $\mathbf{\tau}=\left[  \tau_{11},\tau_{12},...,\tau_{MN}\right]
^{T}$ and the vector of unknowns $\mathbf{\theta}=[x,y,\zeta]^{T}$, we write
(\ref{eq:17}) in vector notation as follows:%

\begin{equation}
\mathbf{\tau}=\mathbf{D\theta},\label{eq:18}%
\end{equation}
where the angle dependent matrix $\mathbf{D}$ is defined as:
\begin{equation}
\mathbf{D}={\normalsize -}\frac{1}{c}\left[
\begin{array}
[c]{ccc}%
a_{tx_{1}}+a_{rx_{1}} & b_{tx_{1}}+b_{rx_{1}} & 1\\
... & ... & ...\\
a_{tx_{M}}+a_{rx_{N}} & b_{tx_{M}}+b_{rx_{N}} & 1
\end{array}
\right]  _{MN\times3}.\label{eq:19}%
\end{equation}
The observation model (\ref{e:mu}) can then be expressed as
\begin{equation}
\mathbf{\mu}=\mathbf{D\theta}+\mathbf{\varepsilon,}\label{eq:20}%
\end{equation}
where $\mathbf{\mu}=\left[  \mu_{11},\mu_{12},...,\mu_{MN}\right]  ^{T},$\ and
$\mathbf{\varepsilon}=\left[  \varepsilon_{11},\varepsilon_{12}%
,...,\varepsilon_{MN}\right]  ^{T}$ is the $MN\times1$ observation noise
vector. To reiterate, a key difference between the MLE and BLUE\ models is
that the MLE target localization is carried out utilizing signal observations
(which are not linear in $x,$ $y)$, while according to (\ref{eq:20}), the
BLUE's \textquotedblleft observations\textquotedblright\ are in the form of
time delays. So an intermediate step of time delay estimation is implied. The
time delays estimates used as observations $\mu_{\ell k}$ can be derived for
example by MLE as follows:%
\begin{equation}
\mu_{\ell k}=\arg\max_{v}\left[  \exp\left(  j2\pi f_{c}v\right)  \int
r_{\ell}\left(  t\right)  s_{k}^{\ast}\left(  t-v\right)  dt\right]
,\label{eq:21}%
\end{equation}
where $v$ is a dummy variable for the time delay.

We still need some characterization of the \textquotedblleft
noise\textquotedblright\ terms $\varepsilon_{\ell k}.$ It is shown in Appendix
\ref{Section:appendixD}, that the maximum likelihood time delay estimates are
unbiased with error covariance matrix
\begin{equation}
\mathbf{C}_{\mathbf{\varepsilon}}=\frac{1}{8\pi^{2}f_{c}^{2}\left\vert
\zeta\right\vert ^{2}/\sigma_{w}^{2}}\mathbf{I}_{MN\times MN},\label{eq:22}%
\end{equation}
where previous definitions of the various quantities apply. For the linear and
Gaussian model in (\ref{eq:20}), the BLUE is computed from the Gauss-Markov
theorem \cite{KaySSPE93} that states the BLUE\ of the unknown vector
$\mathbf{\theta}$ is given by the expression:%

\begin{equation}
\widehat{\mathbf{\theta}}_{B}=\left(  \mathbf{D}^{T}\mathbf{C}%
_{\mathbf{\varepsilon}}^{-1}\mathbf{D}\right)  ^{-1}\mathbf{D}^{T}%
\mathbf{C}_{\mathbf{\varepsilon}}^{-1}\mathbf{\mu}.\label{eq:23}%
\end{equation}
The theorem also establishes that the error covariance matrix is%

\begin{equation}
\mathbf{C}_{B}=\left(  \mathbf{D}^{T}\mathbf{C}_{\mathbf{\varepsilon}}%
^{-1}\mathbf{D}\right)  ^{-1}.\label{eq:CovBLUE}%
\end{equation}

Using the time error covariance matrix $\mathbf{C}_{_{\mathbf{\varepsilon}}} $
and the linear transformation matrix $\mathbf{D}$ in (\ref{eq:19}), the
following estimate for the target localization is obtained:%

\begin{equation}
\left[
\begin{array}
[c]{c}%
\widehat{x}\\
\widehat{y}%
\end{array}
\right]  =\left[  \widehat{\mathbf{\theta}}_{B}\right]  _{2\times1}%
={\small -}c\mathbf{G}_{B}\left[
\begin{array}
[c]{c}%
\overset{M}{\underset{k=1}{\sum}}\overset{N}{\underset{\ell=1}{\sum}}\left(
a_{tx_{k}}+a_{rx_{\ell}}\right)  \mu_{\ell k}\\
\overset{M}{\underset{k=1}{\sum}}\overset{N}{\underset{\ell=1}{\sum}}\left(
b_{tx_{k}}+b_{rx_{\ell}}\right)  \mu_{\ell k}%
\end{array}
\right]  ,\label{eq:BLUE}%
\end{equation}
where $\mu_{\ell k}$ are the time observations, and the matrix $\mathbf{G}%
_{B}$ is of the form:%
\begin{equation}
\mathbf{G}_{B}=\frac{1}{g_{_{1B}}g_{_{2B}}-h_{_{B}}^{2}}\left[
\begin{array}
[c]{cc}%
g_{_{1B}} & h_{_{B}}\\
h_{_{B}} & g_{_{2B}}%
\end{array}
\right]  .
\end{equation}
The elements of matrix $\mathbf{G}_{_{B}}$ are:%

\begin{align}
g_{_{1B}} &  =\overset{M}{\underset{k=1}{\sum}}\overset{N}{\underset{\ell
=1}{\sum}}\left(  b_{tx_{k}}+b_{rx_{\ell}}\right)  ^{2}-\frac{1}{MN}\left(
\overset{M}{\underset{k=1}{\sum}}\overset{N}{\underset{\ell=1}{\sum}}\left(
b_{tx_{k}}+b_{rx_{\ell}}\right)  \right)  ^{2},\label{eq:24}\\
g_{_{2B}} &  =\overset{M}{\underset{k=1}{\sum}}\overset{N}{\underset{\ell
=1}{\sum}}\left(  a_{tx_{k}}+a_{rx_{\ell}}\right)  ^{2}-\frac{1}{MN}\left(
\overset{M}{\underset{k=1}{\sum}}\overset{N}{\underset{\ell=1}{\sum}}\left(
a_{tx_{k}}+a_{rx_{\ell}}\right)  \right)  ^{2},\nonumber\\
h_{_{B}} &  =-\overset{M}{\underset{k=1}{\sum}}\overset{N}{\underset{\ell
=1}{\sum}}\left(  \left(  a_{tx_{k}}+a_{rx_{\ell}}\right)  \left(  b_{tx_{k}%
}+b_{rx_{\ell}}\right)  \right)  ,\nonumber\\
&  +\frac{1}{MN}\overset{M}{\underset{k=1}{\sum}}\overset{N}{\underset{\ell
=1}{\sum}}\left(  a_{tx_{k}}+a_{rx_{\ell}}\right)  \overset{M}{\underset
{k=1}{\sum}}\overset{N}{\underset{\ell=1}{\sum}}\left(  b_{tx_{k}}%
+b_{rx_{\ell}}\right)  .\nonumber
\end{align}
Using these results in (\ref{eq:CovBLUE}) provides the MSE\ for the BLUE as follows:%

\begin{equation}
\sigma_{x,B}^{2}=\frac{c^{2}}{8\pi^{2}f_{c}^{2}\left\vert \zeta\right\vert
^{2}/\sigma_{w}^{2}}\left(  \frac{g_{_{1B}}}{g_{_{1B}}g_{_{2B}}-h_{_{B}}^{2}%
}\right)  ,\label{eq:xMSEblue}%
\end{equation}
for the estimation of the $x$ coordinate, and%

\begin{equation}
\sigma_{y,B}^{2}=\frac{c^{2}}{8\pi^{2}f_{c}^{2}\left\vert \zeta\right\vert
^{2}/\sigma_{w}^{2}}\left(  \frac{g_{_{2B}}}{g_{_{1B}}g_{_{2B}}-h_{_{B}}^{2}%
}\right)  ,\label{eq:yMSEblue}%
\end{equation}
for the estimation of the $y$ coordinate.

\subsection{Discussion}

\label{Section:DiscussionEstimator}

The following points are worth noting:

\begin{itemize}
\item The BLUE estimator in (\ref{eq:23}) and its variance in
(\ref{eq:xMSEblue})\ and (\ref{eq:yMSEblue}) are provided in closed form. This
enables analysis without extensive numerical computations.

\item In general, the variances (\ref{eq:xMSEblue})\ and (\ref{eq:yMSEblue})
have similar functional dependencies on the carrier frequency and on the
sensor deployment as the CRLB (\ref{eq:MSEc}). The terms $a_{tx_{k}},$
$a_{rx_{\ell}},$ $b_{tx_{k}}$ and $b_{rx_{\ell}}$ embedded in
(\ref{eq:xMSEblue})\ and (\ref{eq:yMSEblue}) relate the sensors layout to the
variance of the BLUE .
\end{itemize}

From the expressions of the variance of the BLUE, one can not readily
visualize the effect of the sensors layout. A mapping method, acting as a
design and decision making tool for MIMO radar systems, is proposed and
evaluated in the next subsection.

\subsection{GDOP}

\label{Section:GDOP}

In Section \ref{Section:OptimizationOverAll}, we discussed optimal sensor
location for minimizing the CRLB. In practice, we are faced with a specific
deployment of sensors, and we ask what is the localization accuracy for a
given location of the target. GDOP is a metric that addresses this question.
The GDOP is commonly used in GPS systems for mapping the attainable
localization accuracy for a given layout of GPS satellites positions
\cite{Lee75,LevanonGDOP00}. The GDOP metric emphasizes the effect of sensors
locations by normalizing the localization error with the term contributed by
the range estimate.

The GDOP metric for the two dimensional case is defined:%

\begin{equation}
\text{GDOP}=\frac{\sqrt{\sigma_{x}^{2}+\sigma_{y}^{2}}}{c\sigma_{\varepsilon}%
},\label{eq:GDOPgeneral}%
\end{equation}
where $\sigma_{x}^{2}$\ and $\sigma_{y}^{2}$\ are the variances of
localization on the $x$ and $y$ axis, respectively, and $\sigma_{\varepsilon}$
is the standard deviation of the time delay estimation error, assumed the same
for all sensors. Inherently, the GDOP provides a normalized value that
measures the relative contribution of the radars' location to the overall
accuracy. When the BLUE is used, and the linearity conditions hold,
$\sigma_{x}^{2}$ and $\sigma_{y}^{2}$ are given by (\ref{eq:xMSEblue}) and
(\ref{eq:yMSEblue}), respectively. Using the result in (\ref{eq:22}),
$c\sigma_{\varepsilon}$ for the time delay variance, we get the following GDOP expression:%

\begin{equation}
\text{GDOP}_{B}=\sqrt{\frac{g_{_{1B}}+g_{_{2B}}}{g_{_{1B}}g_{_{2B}}-h_{_{B}%
}^{2}}.}\label{eq:GDOPmimo}%
\end{equation}

The GDOP reduces the combined effect of the locations of the sensors to a
single metric. Once we get the values mapped, the actual localization error is
easily derived by multiplying the GDOP value with $c\sigma_{\varepsilon} $.

\begin{figure}[htp]
\centering\includegraphics[width=11.86 cm,height=9.04 cm]{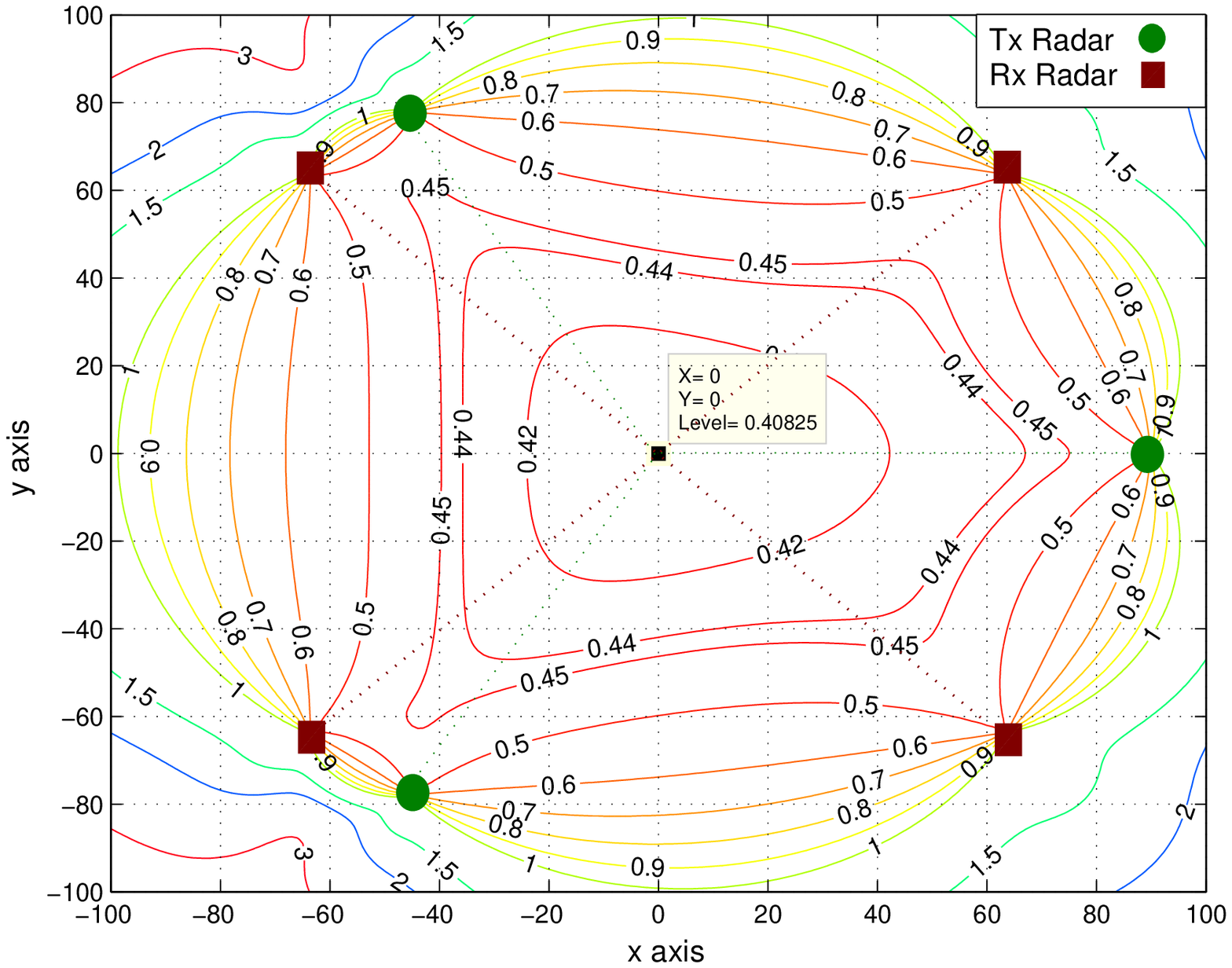}
\caption
{GDOP contours for a symmetric positioning of radars around the axis origin: case (a) with M=3 transmitting radars in a symmetric constellation with the receiving radars set organized in a symmetric constellation of N=4.}%
\label{Fig:2}
\end{figure}

\begin{figure}[htp]
\centering\includegraphics[width=11.86 cm,height=9.04 cm]{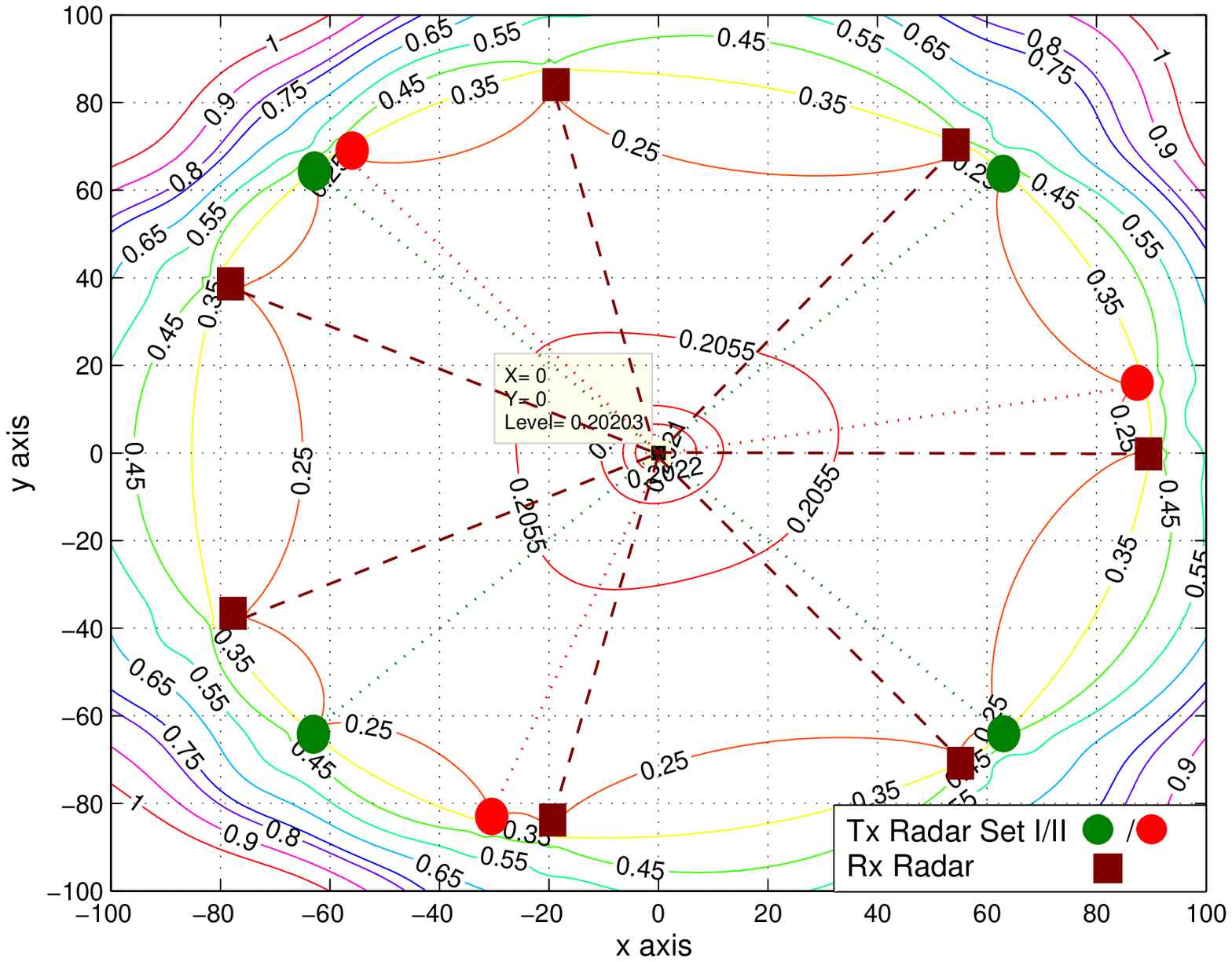}
\caption
{GDOP contours for a symmetric positioning of radars around the axis origin: case (b) with M=7 transmitting radars in two symmetric constellations, one of 3 radars (in red) and second of 4 radars (in green). The receiving radars set are organized in a symmetric constellation of N=7.}%
\label{Fig:3}
\end{figure}

Figure \ref{Fig:2} and \ref{Fig:3} present contour plots of the GDOP values
for $3\times4$ and $7\times7$ MIMO radar systems, respectively. The sensors
are positioned symmetrically around the origin. In Figure \ref{Fig:2}, the
transmitting sensors are located at bearings $\mathbf{\phi}=\left[  \phi
_{i}=\frac{2\pi\left(  i-1\right)  }{3},\ i=1,...,3\right]  ,$ and the
receiving sensors are positioned at bearings $\mathbf{\varphi}=\left[
\varphi_{i}=\frac{\pi}{4}+\frac{2\pi\left(  i-1\right)  }{4}%
,\ i=1,...,4\right]  $. In Figure \ref{Fig:3}, the $M=7$ transmitting sensors
are positioned as a superposition of two symmetrical constellations: the first
set includes three radars and the second four. The sets are located at
bearings $\mathbf{\phi}=\left[  \phi_{i}=\frac{\pi}{18}+\frac{2\pi\left(
i-1\right)  }{3},\ i=1,...,3;\right.  $ $\ \left.  \phi_{i}=\frac{\pi}%
{4}+\frac{2\pi\left(  i-1\right)  }{4},\ \ i=4,...,7\right]  $. The receiving
radars, for this case, are set in a single symmetrical constellation with
bearings $\mathbf{\varphi}=\left[  \varphi_{i}=\frac{2\pi\left(  i-1\right)
}{7},\ i=1,...,7\right]  $. The first noticeable factor in the comparison of
the two plots is the higher accuracy obtained with seven radars compared to
four radars. For example, the lowest GDOP value in Figure \ref{Fig:2}, for the
$3\times4$ system is $0.4082,$ while with seven radars (see Figure
\ref{Fig:3}), the lowest GDOP is $0.2020,$ corresponding to a $50\%$
reduction. When a target is located inside the virtual $\left(  N+M\right)
$-sided system footprint, a higher localization accuracy is obtained than when
a target is outside the footprint of the system. In particular, the best
localization is obtained for a target at the center of the system. The
increase in GDOP values from the center to the footprint boundaries is slow.
Outside the footprint, the GDOP values increase rather rapidly.

\begin{figure}[htp]
\centering\includegraphics[width=11.86 cm,height=9.04 cm]{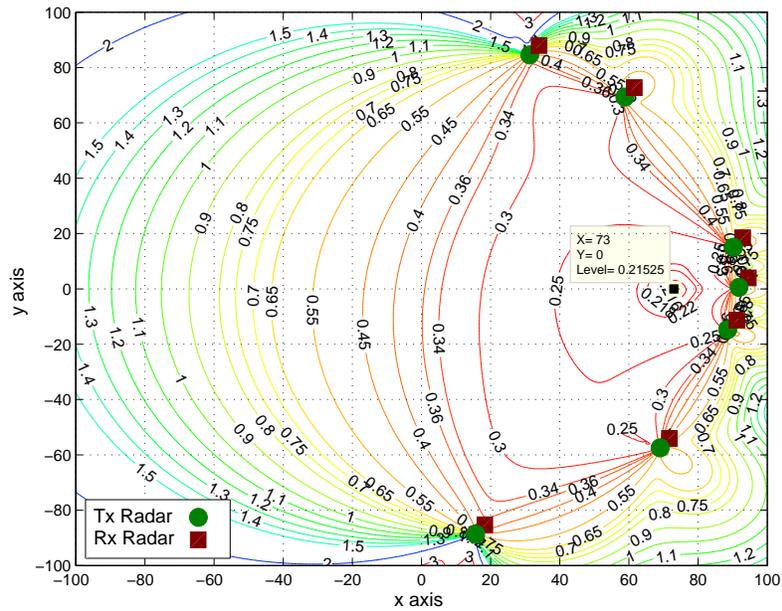}
\caption
{GDOP contours for an asymmetric constellation of the radar set with M=7 transmitting radars and N=7 receiving radars.}%
\label{Fig:4}
\end{figure}

\begin{figure}[htp]
\centering\includegraphics[width=11.86 cm,height=9.04 cm]{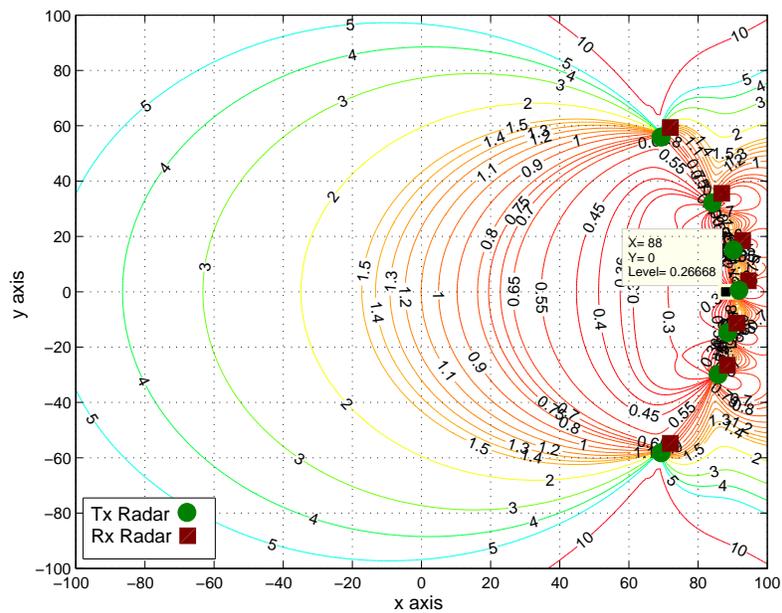}
\caption
{GDOP contours for an asymmetric constellation of the radar set with M=7 transmitting radars and N=7 receiving radars, in the case where the radar are almost aligned.}%
\label{Fig:5}
\end{figure}

In Figure \ref{Fig:4} and Figure \ref{Fig:5}, contours of seven
non-symmetrically positioned radars are drawn. When the radars are relatively
widely spread, as in Figure \ref{Fig:4}, there are still some areas with good
measurement accuracy, though the coverage is shrunk compared to the case with
symmetrical deployment of sensors in Figure \ref{Fig:3}. When the viewing
angle of the target is very restricted, as in Figure \ref{Fig:5}, there is a
marked degradation of GDOP\ values.

These examples demonstrate the main theoretical result of Section IV, namely
that a symmetrical deployment of sensors around the target yields the lowest
GDOP values. Furthermore, calculating the lowest attainable GDOP value using
the optimal results in (\ref{eq:opttrace}) for a $M\times N$ MIMO\ radar, we
obtain a GDOP value of $\sqrt{2/MN}$, and for $M=N$ it is equal to
$\sqrt{2/N^{2}}$. As a numerical example, the lowest GDOPs in Figures
\ref{Fig:2} and \ref{Fig:3} are $\sqrt{2/3\cdot4}\simeq0.4082$ and
$\sqrt{2/7^{2}}\simeq0.2020,\ $respectively. Comparing this with the results
obtained in \cite{LevanonGDOP00} for the case of passive GPS\ based systems,
with $N$ satellites optimally positioned around the target, for which the
lowest achievable GDOP value is $2/\sqrt{N}$, the MIMO\ system advantage is
clearly manifested.

\section{Conclusions}

\label{Section:Conclusions}

In this paper, we have developed analytical expressions for the estimation
errors of coherent and non-coherent MIMO radar using the CRLB. It was shown
that when the processing is coherent and the phase is processed, there is a
reduction in the CRLB values (standard deviation of the estimates) by a factor
of $f_{c}/\beta$ over the case when the observations are non-coherent. We
referred to this gain as coherency gain. Expressions for the CRLB capture also
the impact of the sensors geometry. Further minimization of the localization
error reveals a MIMO radar gain directly proportional to the product of the
number of transmitting and receiving radars. The smallest CRLB is achieved
when the transmitting and receiving sensors are arrayed symmetrically around
the target or any a superposition of such sets. The GDOP metric and mapping
were introduced as a general tool for the analysis of the localization
accuracy with respect to the given radars and target locations. These plots
could serve as a tool for choosing favorable radar locations to cover a given
target area. While localization by coherent MIMO radar provides significantly
better performance than non-coherent processing, it faces the challenge of
multisite systems phase synchronizing, and needs to deal with the ambiguities
stemming from the large separation between sensors.

\appendices%

\section{Derivation of the FIM in (22)}%

\label{Section:appendixA}

In this appendix, we develop the FIM for the unknown parameter vector
$\mathbf{\psi}_{nc}$, based on the conditional pdf in (\ref{eq:pdf_nc}). The
expression for $\mathbf{J}\left(  \mathbf{\psi}\right)  =E\left[
\nabla_{\mathbf{\psi}}\log p\left(  \mathbf{r}|\mathbf{\psi}\right)  \left(
\nabla_{\mathbf{\psi}}\log p\left(  \mathbf{r}|\mathbf{\psi}\right)  \right)
^{H}\right]  =-E\left[  \frac{\partial^{2}\log p\left(  \mathbf{r}%
|\mathbf{\psi}\right)  }{\partial^{2}\mathbf{\psi}}\right]  $ is derived using:%

\begin{align}
&
\begin{array}
[c]{l}%
\left[  \mathbf{J}\left(  \mathbf{\psi}_{nc}\right)  \right]  _{ii^{\prime}%
}=-E\left[  \frac{\partial^{2}\log p\left(  \mathbf{r}|\mathbf{\psi}%
_{nc}\right)  }{\partial\tau_{\ell k}\partial\tau_{\ell^{\prime}k^{\prime}}%
}\right]  ,\\
\left[  \mathbf{J}\left(  \mathbf{\psi}_{nc}\right)  \right]
_{(MN+i),(MN+i^{\prime})}=-E\left[  \frac{\partial^{2}\log p\left(
\mathbf{r}|\mathbf{\psi}_{nc}\right)  }{\partial\alpha_{\ell k}^{R}%
\ \partial\alpha_{\ell^{\prime}k^{\prime}}^{R}}\right]  ,\\
\left[  \mathbf{J}\left(  \mathbf{\psi}_{nc}\right)  \right]
_{(2MN+i),(2MN+i^{\prime})}=-E\left[  \frac{\partial^{2}\log p\left(
\mathbf{r}|\mathbf{\psi}_{nc}\right)  }{\partial\alpha_{\ell k}^{I}%
\ \partial\alpha_{\ell^{\prime}k^{\prime}}^{I}}\right]  ,\\
\left[  \mathbf{J}\left(  \mathbf{\psi}_{nc}\right)  \right]
_{(MN+i),(2MN+i^{\prime})}=\left[  \mathbf{J}\left(  \mathbf{\psi}%
_{nc}\right)  \right]  _{(2MN+i),(MN+i^{\prime})}=-E\left[  \frac{\partial
^{2}\log p\left(  \mathbf{r}|\mathbf{\psi}_{nc}\right)  }{\partial\alpha_{\ell
k}^{R}\ \partial\alpha_{\ell^{\prime}k^{\prime}}^{I}}\right]  ,\\
\left[  \mathbf{J}\left(  \mathbf{\psi}_{nc}\right)  \right]  _{i,\left(
MN+i^{\prime}\right)  }=\left[  \mathbf{J}\left(  \mathbf{\psi}_{nc}\right)
\right]  _{\left(  MN+i\right)  ,i^{\prime}}=-E\left[  \frac{\partial^{2}\log
p\left(  \mathbf{r}|\mathbf{\psi}_{nc}\right)  }{\partial\tau_{\ell
k}\ \partial\alpha_{\ell k}^{R}}\right]  ,\\
\left[  \mathbf{J}\left(  \mathbf{\psi}_{nc}\right)  \right]  _{i,\left(
2MN+i^{\prime}\right)  }=\left[  \mathbf{J}\left(  \mathbf{\psi}_{nc}\right)
\right]  _{\left(  2MN+i\right)  ,i^{\prime}}=-E\left[  \frac{\partial^{2}\log
p\left(  \mathbf{r}|\mathbf{\psi}_{nc}\right)  }{\partial\tau_{\ell
k}\ \partial\alpha_{\ell k}^{I}}\right]  ,
\end{array}
\label{eq:App_a1}\\
&
\begin{array}
[c]{cc}%
i=(\ell-1)\ast M+k, & i^{\prime}=(\ell^{\prime}-1)\ast M+k^{\prime},\\
\ell,\ell^{\prime}=1,..,N; & k,k^{\prime}=1,..,M;
\end{array}
\text{ \ \ }\nonumber
\end{align}
The first derivative of $p\left(  \mathbf{r}|\mathbf{\psi}_{nc}\right)  $ with
respect to the elements of $\mathbf{\tau}$ is:%
\begin{align}
\frac{\partial\left[  \log p\left(  \mathbf{r}|\mathbf{\psi}_{nc}\right)
\right]  }{\partial\tau_{\ell k}}= &  \text{$\frac{1}{\sigma_{w}^{2}}$}%
{\textstyle\int}
\left\{  \left[  {\small r}_{\ell}{\small (t)-}\overset{M}{\underset
{k^{\prime}=1}{\sum}}\alpha_{\ell k^{\prime}}{\small s}_{k^{\prime}}\left(
t-\tau_{\ell k^{\prime}}\right)  \right]  \cdot\alpha_{\ell k}^{\ast}%
\frac{{\small \partial}\left[  {\small s}_{k}^{\ast}\left(  t-\tau_{\ell
k}\right)  \right]  }{{\small \partial\tau}_{\ell k}}\right. \label{eq:App_a3}%
\\
&  \left.  +\left[  {\small r}_{\ell}{\small (t)-}\overset{M}{\underset
{k^{\prime}=1}{\sum}}\alpha_{\ell k^{\prime}}{\small s}_{k^{\prime}}\left(
t-\tau_{\ell k^{\prime}}\right)  \right]  ^{\ast}\cdot\alpha_{\ell k}%
\frac{{\small \partial}\left[  {\small s}_{k}\left(  t-\tau_{\ell k}\right)
\right]  }{{\small \partial\tau}_{\ell k}}\right\}  dt.\nonumber
\end{align}
Applying the second derivative to (\ref{eq:App_a3}), define a matrix
$\mathbf{S}_{nc}$ with the following elements:%
\begin{align}
\lbrack\mathbf{S}_{nc}]_{ii^{\prime}} &  =\frac{\sigma_{w}^{2}}{2}%
[{\small \mathbf{J}\left(  \mathbf{\psi}\right)  }]_{ii^{\prime}%
}=\label{eq:App_a4}\\
&  =E\left\{  \frac{\partial^{2}}{\partial{\tau_{\ell k}}\partial{\tau
_{\ell^{\prime}k^{\prime}}}}\int\left[  \alpha_{\ell k}{\small s}_{k}\left(
t-{\tau_{\ell k}}\right)  \alpha_{\ell k^{\prime}}^{\ast}{\small s}%
_{k^{\prime}}^{\ast}\left(  t-{\tau_{\ell k^{\prime}}}\right)  \right.
\right. \nonumber\\
&  \left.  +\left.  \alpha_{\ell k}^{\ast}{\small s}_{k}^{\ast}\left(
t-{\tau_{\ell k}}\right)  \alpha_{\ell k^{\prime}}{\small s}_{k^{\prime}%
}\left(  t-{\tau_{\ell k^{\prime}}}\right)  \right]  dt\right\} \nonumber\\
&  =\operatorname{Re}\left\{  \alpha_{\ell k}\alpha_{\ell^{\prime}k^{\prime}%
}^{\ast}\left[  \frac{\partial^{2}}{\partial{\tau_{\ell k}}\partial{\tau
_{\ell^{\prime}k^{\prime}}}}\int{\small s}_{k}\left(  t-{\tau_{\ell k}%
}\right)  {\small s}_{k^{\prime}}^{\ast}\left(  t-{\tau_{\ell k^{\prime}}%
}\right)  dt\right]  \right\}  .\nonumber
\end{align}

Using matrix notation for compactness,
\begin{equation}
\mathbf{S}_{nc}=\frac{\partial^{2}}{\partial\mathbf{\tau}^{2}}%
\operatorname{Re}\left[  \operatorname*{diag}(\mathbf{\alpha})\mathbf{R}%
_{s}\operatorname*{diag}\left(  \mathbf{\alpha}^{\ast}\right)  \right]
,\label{e:Snc}%
\end{equation}
where $\operatorname*{diag}(\cdot)$ denotes a diagonal matrix, $\mathbf{\alpha
}$ was defined in (\ref{eq:alpha_nc}), and we abuse the notation and let%

\begin{equation}
\left[  \frac{\partial^{2}}{\partial\mathbf{\tau}^{2}}\mathbf{R}_{s}\right]
_{ii^{\prime}}\equiv\frac{\partial}{\partial{\tau_{\ell k}}\partial{\tau
_{\ell^{\prime}k^{\prime}}}}\left[  \mathbf{R}_{s}\right]  _{ii^{\prime}%
}.\label{e:d_sqr_Rs}%
\end{equation}
The elements of matrix $\mathbf{R}_{s}$ are defined as:%

\begin{equation}
\left[  \mathbf{R}_{s}\right]  _{ii^{\prime}}\equiv\left\{
\begin{array}
[c]{cc}%
\int{\small s}_{k}\left(  t-{\tau_{\ell k}}\right)  {\small s}_{k^{\prime}%
}^{\ast}\left(  t-{\tau_{\ell k^{\prime}}}\right)  dt & \ell=\ell^{\prime}\\
0 & \ell\neq\ell^{\prime}%
\end{array}
\right.  .\label{e:Rs}%
\end{equation}

The second and third terms in (\ref{eq:App_a1}) define a matrix
$\mathbf{\Lambda}_{\alpha}$ with the following elements:%
\begin{align}
\lbrack\mathbf{\Lambda}_{\alpha}]_{ii^{\prime}} &  =[\mathbf{\Lambda}_{\alpha
}]_{\left(  MN+i\right)  ,\left(  MN+i^{\prime}\right)  }=\frac{\sigma_{w}%
^{2}}{2}\left[  \mathbf{J}\left(  \mathbf{\psi}_{nc}\right)  \right]
_{(MN+i),(MN+i^{\prime})}=\frac{\sigma_{w}^{2}}{2}\left[  \mathbf{J}\left(
\mathbf{\psi}_{nc}\right)  \right]  _{(2MN+i),(2MN+i^{\prime})}%
\label{eq:App_a5}\\
&  =E\left\{  \frac{\partial}{\partial\alpha_{\ell^{\prime}k^{\prime}}^{R}%
}\int\left[  \overset{M}{\underset{k^{\prime}=1}{\sum}}{\small s}_{k}\left(
t-{\tau_{\ell k}}\right)  \alpha_{\ell k^{\prime}}^{\ast}{\small s}%
_{k^{\prime}}^{\ast}\left(  t-{\tau_{\ell k^{\prime}}}\right)  \right.
\right. \nonumber\\
&  \left.  \left.  +\overset{M}{\underset{k^{\prime}=1}{\sum}}{\small s}%
_{k}^{\ast}\left(  t-{\tau_{\ell k}}\right)  \alpha_{\ell k^{\prime}%
}{\small s}_{k^{\prime}}\left(  t-{\tau_{\ell k^{\prime}}}\right)  \right]
dt\right\} \nonumber\\
&  =\operatorname{Re}\left\{  \left[  \mathbf{R}_{s}\right]  _{ii^{\prime}%
}\right\}  ,\nonumber
\end{align}
and%

\begin{align}
\lbrack\mathbf{\Lambda}_{\alpha}]_{i,\left(  MN+i^{\prime}\right)  } &
=[\mathbf{\Lambda}_{\alpha}]_{\left(  MN+i\right)  ,i^{\prime}}=\frac
{\sigma_{w}^{2}}{2}\left[  \mathbf{J}\left(  \mathbf{\psi}_{nc}\right)
\right]  _{(MN+i),(2MN+i^{\prime})}=\frac{\sigma_{w}^{2}}{2}\left[
\mathbf{J}\left(  \mathbf{\psi}_{nc}\right)  \right]  _{(2MN+i),(MN+i^{\prime
})}\label{eq:App_a6}\\
&  =E\left\{  \frac{\partial}{\partial\alpha_{\ell^{\prime}k^{\prime}}^{I}%
}\int\left[  \overset{M}{\underset{k^{\prime}=1}{\sum}}\left(  j\right)
{\small s}_{k}\left(  t-{\tau_{\ell k}}\right)  \alpha_{\ell k^{\prime}}%
^{\ast}{\small s}_{k^{\prime}}^{\ast}\left(  t-{\tau_{\ell k^{\prime}}%
}\right)  \right.  \right. \nonumber\\
&  \left.  \left.  +\overset{M}{\underset{k^{\prime}=1}{\sum}}\left(
-j\right)  {\small s}_{k}^{\ast}\left(  t-{\tau_{\ell k}}\right)  \alpha_{\ell
k^{\prime}}{\small s}_{k^{\prime}}\left(  t-{\tau_{\ell k^{\prime}}}\right)
\right]  dt\right\} \nonumber\\
&  =-\operatorname{Im}\left\{  \left[  \mathbf{R}_{s}\right]  _{ii^{\prime}%
}\right\}  .\nonumber
\end{align}
In matrix notation,
\begin{equation}
\mathbf{\Lambda}_{\alpha}=\left[
\begin{array}
[c]{cc}%
\operatorname{Re}\left[  \mathbf{R}_{s}\right]  & -\operatorname{Im}\left[
\mathbf{R}_{s}\right] \\
-\operatorname{Im}\left[  \mathbf{R}_{s}\right]  & \operatorname{Re}\left[
\mathbf{R}_{s}\right]
\end{array}
\right]  .\label{e:Lambdanc}%
\end{equation}

The fourth and fifth terms in (\ref{eq:App_a1}) define the matrix
$\mathbf{V}_{nc}$ with the following elements:%
\begin{align}
\lbrack\mathbf{V}_{nc}]_{ii^{\prime}} &  =\frac{\sigma_{w}^{2}}{2}\left[
\mathbf{J}\left(  \mathbf{\psi}_{nc}\right)  \right]  _{(MN+i),i^{\prime}%
}=\frac{\sigma_{w}^{2}}{2}\left[  \mathbf{J}\left(  \mathbf{\psi}_{nc}\right)
\right]  _{i,(MN+i^{\prime})}\label{eq:App_a7}\\
&  =E\left\{  \frac{{\small \partial}}{{\small \partial\tau}_{\ell k}}%
\frac{\partial}{\partial\alpha^{R}{_{\ell^{\prime}k^{\prime}}}}\int\left[
\alpha_{\ell k}{\small s}_{k}\left(  t-\tau_{\ell k}\right)  \alpha_{\ell
k^{\prime}}^{\ast}{\small s}_{k^{\prime}}^{\ast}\left(  t-\tau_{\ell
k^{\prime}}\right)  \right.  \right. \nonumber\\
&  \left.  \left.  +\alpha_{\ell k}^{\ast}{\small s}_{k}^{\ast}\left(
t-\tau_{\ell k}\right)  \alpha_{\ell k^{\prime}}{\small s}_{k^{\prime}}\left(
t-\tau_{\ell k^{\prime}}\right)  \right]  dt\right\} \nonumber\\
&  =\operatorname{Re}\left\{  \alpha_{\ell k}\frac{{\small \partial}%
}{{\small \partial\tau}_{\ell k}}\left[  \mathbf{R}_{s}\right]  _{ii^{\prime}%
}\right\}  ,\nonumber
\end{align}
and%
\begin{align}
\lbrack\mathbf{V}_{nc}]_{i,\left(  MN+i^{\prime}\right)  } &  =\frac
{\sigma_{w}^{2}}{2}\left[  \mathbf{J}\left(  \mathbf{\psi}_{nc}\right)
\right]  _{(2MN+i),i^{\prime}}=\frac{\sigma_{w}^{2}}{2}\left[  \mathbf{J}%
\left(  \mathbf{\psi}_{nc}\right)  \right]  _{i,(2MN+i^{\prime})}%
\label{eq:App_a8}\\
&  =E\left\{  \frac{{\small \partial}}{{\small \partial\tau}_{\ell k}}%
\frac{\partial}{\partial\alpha^{I}{_{\ell^{\prime}k^{\prime}}}}\int\left[
\alpha_{\ell k}{\small s}_{k}\left(  t-\tau_{\ell k}\right)  \alpha_{\ell
k^{\prime}}^{\ast}{\small s}_{k^{\prime}}^{\ast}\left(  t-\tau_{\ell
k^{\prime}}\right)  \right.  \right. \nonumber\\
&  \left.  \left.  +\alpha_{\ell k}^{\ast}{\small s}_{k}^{\ast}\left(
t-\tau_{\ell k}\right)  \alpha_{\ell k^{\prime}}{\small s}_{k^{\prime}}\left(
t-\tau_{\ell k^{\prime}}\right)  \right]  dt\right\} \nonumber\\
&  =-\operatorname{Im}\left\{  \alpha_{\ell k}\frac{{\small \partial}%
}{{\small \partial\tau}_{\ell k}}\left[  \mathbf{R}_{s}\right]  _{ii^{\prime}%
}\right\}  .\nonumber
\end{align}
In matrix notation:
\begin{equation}
\mathbf{V}_{nc}=\left[
\begin{array}
[c]{cc}%
\frac{\partial}{\partial\mathbf{\tau}}\operatorname{Re}\left[
\operatorname*{diag}(\mathbf{\alpha})\mathbf{R}_{s}\right]  ; & -\frac
{\partial}{\partial\mathbf{\tau}}\operatorname{Im}\left[  \operatorname*{diag}%
(\mathbf{\alpha})\mathbf{R}_{s}\right]
\end{array}
\right]  .\label{e:Vnc}%
\end{equation}
\bigskip

\emph{Orthogonal Waveforms }

Orthogonality implies that all cross elements $\int{\small s}_{k}\left(
t-\tau_{\ell k}\right)  {\small s}_{k^{\prime}}^{\ast}\left(  t-\tau
_{\ell^{\prime}k^{\prime}}\right)  dt=0,$ for $\ell\neq\ell^{\prime}$ and
$k\neq k^{\prime},$and after some algebra, the matrices defined by
(\ref{eq:App_a4})-(\ref{eq:App_a8}) take the following form:%
\begin{equation}%
\begin{array}
[c]{c}%
\left[  \mathbf{S}_{nc}\right]  _{ii^{\prime}}=\left\{
\begin{array}
[c]{cc}%
4\pi^{2}\beta^{2}\left[  \left\vert \alpha_{lk}\right\vert ^{2}\beta_{R_{k}%
}^{2}\right]  & i=i^{\prime}\\
0 & i\neq i^{\prime}%
\end{array}
\right. \\
\lbrack\mathbf{\Lambda}_{\alpha}]_{ii^{\prime}}=[\mathbf{\Lambda}_{\alpha
}]_{\left(  MN+i\right)  ,\left(  MN+i^{\prime}\right)  }=\left\{
\begin{array}
[c]{cc}%
1 & i=i^{\prime}\\
0 & i\neq i^{\prime}%
\end{array}
\right. \\
\lbrack\mathbf{\Lambda}_{\alpha}]_{i,\left(  MN+i^{\prime}\right)
}=[\mathbf{\Lambda}_{\alpha}]_{\left(  MN+i\right)  ,i^{\prime}}=0\\
\lbrack\mathbf{V}_{nc}]_{ii^{\prime}}=0\\
\lbrack\mathbf{V}_{nc}]_{i,\left(  MN+i^{\prime}\right)  }=0.
\end{array}
\label{eq:App_a9}%
\end{equation}
\bigskip%

\section{Derivation of the FIM in (34)}%

\label{Section:appendixB}

In this appendix, we develop the FIM for the unknown parameter vector
$\mathbf{\psi}_{c}$, based on the conditional pdf in (\ref{eq:pdf_c}). The
expression for $\mathbf{J}\left(  \mathbf{\psi}\right)  =E\left\{
\nabla_{\mathbf{\psi}}\log p\left(  \mathbf{r}|\mathbf{\psi}\right)  \left(
\nabla_{\mathbf{\psi}}\log p\left(  \mathbf{r}|\mathbf{\psi}\right)  \right)
^{H}\right\}  =-E\left[  \frac{\partial^{2}\log p\left(  \mathbf{r}%
|\mathbf{\psi}\right)  }{\partial^{2}\mathbf{\psi}}\right]  $ is derived using:%

\begin{align}
&
\begin{array}
[c]{l}%
\left[  \mathbf{J}\left(  \mathbf{\psi}_{c}\right)  \right]  _{ii^{\prime}%
}=-E\left[  \frac{\partial^{2}\log p\left(  \mathbf{r}|\mathbf{\psi}%
_{c}\right)  }{\partial\tau_{\ell k}\partial\tau_{\ell^{\prime}k^{\prime}}%
}\right]  ,\\
\left[  \mathbf{J}\left(  \mathbf{\psi}_{c}\right)  \right]  _{(MN+1),(MN+1)}%
=-E\left[  \frac{\partial^{2}\log p\left(  \mathbf{r}|\mathbf{\psi}%
_{c}\right)  }{\left(  \partial\zeta^{R}\right)  ^{2}\ }\right]  ,\\
\left[  \mathbf{J}\left(  \mathbf{\psi}_{c}\right)  \right]  _{(MN+2),(MN+2)}%
=-E\left[  \frac{\partial^{2}\log p\left(  \mathbf{r}|\mathbf{\psi}%
_{c}\right)  }{\left(  \partial\zeta^{I}\right)  ^{2}}\right]  ,\\
\left[  \mathbf{J}\left(  \mathbf{\psi}_{c}\right)  \right]  _{(MN+1),(MN+2)}%
=\left[  \mathbf{J}\left(  \mathbf{\psi}_{c}\right)  \right]  _{(MN+2),(MN+1)}%
=-E\left[  \frac{\partial^{2}\log p\left(  \mathbf{r}|\mathbf{\psi}%
_{c}\right)  }{\partial\zeta^{R}\ \partial\zeta^{I}}\right]  ,\\
\left[  \mathbf{J}\left(  \mathbf{\psi}_{c}\right)  \right]  _{i,\left(
MN+1\right)  }=\left[  \mathbf{J}\left(  \mathbf{\psi}_{c}\right)  \right]
_{\left(  MN+1\right)  ,i^{\prime}}=-E\left[  \frac{\partial^{2}\log p\left(
\mathbf{r}|\mathbf{\psi}_{c}\right)  }{\partial\tau_{\ell k}\ \partial
\zeta^{R}\ }\right]  ,\\
\left[  \mathbf{J}\left(  \mathbf{\psi}_{c}\right)  \right]  _{i,\left(
MN+2\right)  }=\left[  \mathbf{J}\left(  \mathbf{\psi}_{c}\right)  \right]
_{\left(  MN+2\right)  ,i^{\prime}}=-E\left[  \frac{\partial^{2}\log p\left(
\mathbf{r}|\mathbf{\psi}_{c}\right)  }{\partial\tau_{\ell k}\ \partial
\zeta^{I}}\right]  ,
\end{array}
\label{eq:App_b1}\\
&
\begin{array}
[c]{cc}%
i=(\ell-1)\ast M+k, & i^{\prime}=(\ell^{\prime}-1)\ast M+k^{\prime},\\
\ell,\ell^{\prime}=1,..,N; & k,k^{\prime}=1,..,M.
\end{array}
\text{ \ .\ }\nonumber
\end{align}
The first derivative of $p\left(  \mathbf{r}|\mathbf{\psi}_{c}\right)  $ with
respect to the elements of $\mathbf{\tau}$ is:%
\begin{align}
\frac{\partial\left[  \log p\left(  \mathbf{r}|\mathbf{\psi}_{c}\right)
\right]  }{\partial\tau_{\ell k}}= &  \text{$\frac{1}{\sigma_{w}^{2}}$}%
{\textstyle\int}
\left\{  \left[  {\small r}_{\ell}{\small (t)-}\overset{M}{\underset
{k^{\prime}=1}{\sum}}\zeta\exp\left(  -j2\pi f_{c}\tau_{\ell k^{\prime}%
}\right)  {\small s}_{k^{\prime}}\left(  t-\tau_{\ell k^{\prime}}\right)
\right]  \cdot\zeta^{\ast}\frac{{\small \partial}\left[  \exp\left(  j2\pi
f_{c}\tau_{\ell k}\right)  {\small s}_{k}^{\ast}\left(  t-\tau_{\ell
k}\right)  \right]  }{{\small \partial\tau}_{\ell k}}\right. \label{eq:App_b3}%
\\
&  \left.  +\left[  {\small r}_{\ell}{\small (t)-}\overset{M}{\underset
{k^{\prime}=1}{\sum}}\zeta\exp\left(  -j2\pi f_{c}\tau_{\ell k^{\prime}%
}\right)  {\small s}_{k^{\prime}}\left(  t-\tau_{\ell k^{\prime}}\right)
\right]  ^{\ast}\cdot\zeta\frac{{\small \partial}\left[  \exp\left(  -j2\pi
f_{c}\tau_{\ell k}\right)  {\small s}_{k}\left(  t-\tau_{\ell k}\right)
\right]  }{{\small \partial\tau}_{\ell k}}\right\}  dt.\nonumber
\end{align}
Applying the second derivative to (\ref{eq:App_b3}) define a matrix
$\mathbf{S}_{nc}$ with the following elements:%
\begin{align}
\lbrack\mathbf{S}_{c}]_{ii^{\prime}} &  =\frac{\sigma_{w}^{2}}{2}%
[{\small \mathbf{J}\left(  \mathbf{\psi}\right)  }]_{ii^{\prime}%
}=\label{eq:App_b4}\\
&  =E\left\{  \frac{\partial^{2}}{\partial{\tau_{\ell k}}\partial{\tau
_{\ell^{\prime}k^{\prime}}}}\int\left[  \zeta\zeta^{\ast}\exp\left(  j2\pi
f_{c}\left(  \tau_{\ell k}-\tau_{\ell^{\prime}k^{\prime}}\right)  \right)
{\small s}_{k^{\prime}}\left(  t-{\tau_{\ell k^{\prime}}}\right)
{\small s}_{k}^{\ast}\left(  t-{\tau_{\ell k}}\right)  \right.  \right.
\nonumber\\
&  \left.  +\left.  \zeta^{\ast}\zeta\exp\left(  -j2\pi\left(  \tau_{\ell
k}-\tau_{\ell k^{\prime}}\right)  \right)  {\small s}_{k^{\prime}}^{\ast
}\left(  t-{\tau_{\ell k^{\prime}}}\right)  {\small s}_{k}\left(
t-{\tau_{\ell k}}\right)  \right]  dt\right\} \nonumber\\
&  =\operatorname{Re}\left\{  \left\vert \zeta\right\vert ^{2}\left[
\frac{\partial^{2}}{\partial{\tau_{\ell k}}\partial{\tau_{\ell^{\prime
}k^{\prime}}}}\left(  \exp\left(  -j2\pi f_{c}\left(  \tau_{\ell k}-\tau
_{\ell^{\prime}k^{\prime}}\right)  \right)  \left[  \mathbf{R}_{s}\right]
_{ii^{\prime}}\right)  \right]  \right\}  .\nonumber
\end{align}
In matrix form,
\begin{equation}
\mathbf{S}_{c}=\left\vert \zeta\right\vert ^{2}\frac{\partial^{2}}%
{\partial\mathbf{\tau}^{2}}\operatorname{Re}\left\{  \operatorname*{diag}%
(\mathbf{e)R}_{s}\operatorname*{diag}\left(  \mathbf{e}^{^{\ast}}\right)
\right\}  ,\label{e:Sc}%
\end{equation}
where the operator $\frac{\partial^{2}}{\partial\mathbf{\tau}^{2}}$ and the
matrix $\mathbf{\mathbf{R}_{s}}$ were defined in Appendix
\ref{Section:appendixA}, $\mathbf{e}=\left[  \exp\left(  -2\pi f_{c}\tau
_{11}\right)  ,\exp\left(  -2\pi f_{c}\tau_{12}\right)  ,\right.  $ $\left.
...,\exp\left(  -2\pi f_{c}\tau_{MN}\right)  \right]  $.

The second and third terms in (\ref{eq:App_b1}) define a matrix
$\mathbf{\Lambda}_{\alpha c}$ with the following elements:%
\begin{align}
\lbrack\mathbf{\Lambda}_{\alpha c}]_{11} &  =[\mathbf{\Lambda}_{\alpha
c}]_{22}=\frac{\sigma_{w}^{2}}{2}\left[  \mathbf{J}\left(  \mathbf{\psi}%
_{c}\right)  \right]  _{(MN+1),(MN+1)}=\frac{\sigma_{w}^{2}}{2}\left[
\mathbf{J}\left(  \mathbf{\psi}_{c}\right)  \right]  _{(MN+2),(MN+2)}%
\label{eq:App_b6}\\
&  =E\left\{  \overset{N}{\underset{\ell=1}{\sum}}\overset{M}{\underset
{k=1}{\sum}}\int\left[  \overset{M}{\underset{k^{\prime}=1}{\sum}}\exp\left(
-j2\pi f_{c}\left(  \tau_{\ell k}-\tau_{\ell k^{\prime}}\right)  \right)
{\small s}_{k}\left(  t-{\tau_{\ell k}}\right)  {\small s}_{k^{\prime}}^{\ast
}\left(  t-{\tau_{\ell k^{\prime}}}\right)  \right.  \right. \nonumber\\
&  \left.  \left.  +\overset{M}{\underset{k^{\prime}=1}{\sum}}\exp\left(
j2\pi f_{c}\left(  \tau_{\ell k}-\tau_{\ell k^{\prime}}\right)  \right)
{\small s}_{k}^{\ast}\left(  t-{\tau_{\ell k}}\right)  {\small s}_{k^{\prime}%
}\left(  t-{\tau_{\ell k^{\prime}}}\right)  \right]  dt\right\} \nonumber\\
&  =\operatorname{Re}\left\{  \overset{N}{\underset{\ell=1}{\sum}}\overset
{N}{\underset{\ell^{\prime}=1}{\sum}}\overset{M}{\underset{k=1}{\sum}}%
\overset{M}{\underset{k^{\prime}=1}{\sum}}\exp\left(  -j2\pi f_{c}\left(
\tau_{\ell k}-\tau_{\ell^{\prime}k^{\prime}}\right)  \right)  \left[
\mathbf{R}_{s}\right]  _{ii^{\prime}}\right\}  ,\nonumber
\end{align}
and%

\begin{align}
\lbrack\mathbf{\Lambda}_{\alpha c}]_{12} &  =[\mathbf{\Lambda}_{\alpha
c}]_{21}=\frac{\sigma_{w}^{2}}{2}\left[  \mathbf{J}\left(  \mathbf{\psi}%
_{c}\right)  \right]  _{(MN+1)(MN+2)}=\frac{\sigma_{w}^{2}}{2}\left[
\mathbf{J}\left(  \mathbf{\psi}_{c}\right)  \right]  _{(MN+2)(MN+1)}%
=\label{eq:App_b7}\\
&  =E\left\{  \overset{N}{\underset{\ell=1}{\sum}}\overset{M}{\underset
{k=1}{\sum}}\int\left[  \overset{M}{\underset{k^{\prime}=1}{\sum}}\left(
j\right)  ^{\ast}\exp\left(  -j2\pi f_{c}\left(  \tau_{\ell k}-\tau_{\ell
k^{\prime}}\right)  \right)  {\small s}_{k}\left(  t-{\tau_{\ell k}}\right)
{\small s}_{k^{\prime}}^{\ast}\left(  t-{\tau_{\ell k^{\prime}}}\right)
\right.  \right. \nonumber\\
&  \left.  \left.  +\overset{M}{\underset{k^{\prime}=1}{\sum}}\left(
j\right)  \exp\left(  j2\pi f_{c}\left(  \tau_{\ell k}-\tau_{\ell k^{\prime}%
}\right)  \right)  {\small s}_{k}^{\ast}\left(  t-{\tau_{\ell k}}\right)
{\small s}_{k^{\prime}}\left(  t-{\tau_{\ell k^{\prime}}}\right)  \right]
dt\right\} \nonumber\\
&  =-\operatorname{Im}\left\{  \overset{N}{\underset{\ell=1}{\sum}}\overset
{N}{\underset{\ell^{\prime}=1}{\sum}}\overset{M}{\underset{k=1}{\sum}}%
\overset{M}{\underset{k^{\prime}=1}{\sum}}\exp\left(  -j2\pi f_{c}\left(
\tau_{\ell k}-\tau_{\ell^{\prime}k^{\prime}}\right)  \right)  \left[
\mathbf{R}_{s}\right]  _{ii^{\prime}}\right\}  .\nonumber
\end{align}
In matrix form,
\begin{equation}
\mathbf{\Lambda}_{\alpha c}=\left[
\begin{array}
[c]{cc}%
\operatorname{Re}\left[  \mathbf{eR}_{s}\mathbf{e}^{H}\right]  &
-\operatorname{Im}\left[  \mathbf{eR}_{s}\mathbf{e}^{H}\right] \\
-\operatorname{Im}\left[  \mathbf{eR}_{s}\mathbf{e}^{H}\right]  &
\operatorname{Re}\left[  \mathbf{eR}_{s}\mathbf{e}^{H}\right]
\end{array}
\right]  .\label{e:Lambdac}%
\end{equation}

The fourth and fifth terms in (\ref{eq:App_b1}) define the matrix
$\mathbf{V}_{c}$ with the following elements:%
\begin{align}
\lbrack\mathbf{V}_{c}]_{i1} &  =\frac{\sigma_{w}^{2}}{2}\left[  \mathbf{J}%
\left(  \mathbf{\psi}_{c}\right)  \right]  _{i,(MN+1)}=\frac{\sigma_{w}^{2}%
}{2}\left[  \mathbf{J}\left(  \mathbf{\psi}_{c}\right)  \right]
_{(MN+1),i^{\prime}}\label{eq:App_b8}\\
&  =E\left\{  \frac{\partial}{\partial{\tau_{\ell k}}}\int\left[
\zeta\overset{M}{\underset{k^{\prime}=1}{\sum}}\exp\left(  -j2\pi f_{c}\left(
\tau_{\ell k}-\tau_{\ell k^{\prime}}\right)  \right)  {\small s}_{k}\left(
t-\tau_{\ell k}\right)  {\small s}_{k^{\prime}}^{\ast}\left(  t-\tau_{\ell
k^{\prime}}\right)  \right.  \right. \nonumber\\
&  \left.  \left.  +\zeta^{\ast}\overset{M}{\underset{k^{\prime}=1}{\sum}}%
\exp\left(  j2\pi f_{c}\left(  \tau_{\ell k}-\tau_{\ell k^{\prime}}\right)
\right)  {\small s}_{k}^{\ast}\left(  t-\tau_{\ell k}\right)  {\small s}%
_{k^{\prime}}\left(  t-\tau_{\ell k^{\prime}}\right)  \right]  dt\right\}
\nonumber\\
&  =\frac{{\small \partial}}{{\small \partial\tau}_{\ell k}}\operatorname{Re}%
\left\{  \overset{N}{\underset{\ell^{\prime}=1}{\sum}}\overset{M}%
{\underset{k^{\prime}=1}{\sum}}\zeta\exp\left(  -j2\pi f_{c}\left(  \tau_{\ell
k}-\tau_{\ell^{\prime}k^{\prime}}\right)  \right)  \left[  \mathbf{R}%
_{s}\right]  _{ii^{\prime}}dt\right\}  ,\nonumber
\end{align}
and%
\begin{align}
\lbrack\mathbf{V}_{c}]_{i2} &  =\frac{\sigma_{w}^{2}}{2}\left[  \mathbf{J}%
\left(  \mathbf{\psi}_{c}\right)  \right]  _{i,(MN+2)}=\frac{\sigma_{w}^{2}%
}{2}\left[  \mathbf{J}\left(  \mathbf{\psi}_{c}\right)  \right]
_{(MN+2),i^{\prime}}\label{eq:App_b9}\\
&  =E\left\{  \frac{\partial}{\partial{\tau_{\ell k}}}\int\left[  \left(
j\zeta\right)  \overset{M}{\underset{k^{\prime}=1}{\sum}}\exp\left(  -j2\pi
f_{c}\left(  \tau_{\ell k}-\tau_{\ell k^{\prime}}\right)  \right)
{\small s}_{k}\left(  t-\tau_{\ell k}\right)  {\small s}_{k^{\prime}}^{\ast
}\left(  t-\tau_{\ell k^{\prime}}\right)  \right.  \right. \nonumber\\
&  \left.  \left.  +\left(  j\zeta\right)  ^{\ast}\overset{M}{\underset
{k^{\prime}=1}{\sum}}\exp\left(  j2\pi f_{c}\left(  \tau_{\ell k}-\tau_{\ell
k^{\prime}}\right)  \right)  {\small s}_{k}^{\ast}\left(  t-\tau_{\ell
k}\right)  {\small s}_{k^{\prime}}\left(  t-\tau_{\ell k^{\prime}}\right)
\right]  dt\right\} \nonumber\\
&  =-\frac{{\small \partial}}{{\small \partial\tau}_{\ell k}}\operatorname{Im}%
\left\{  \overset{N}{\underset{\ell^{\prime}=1}{\sum}}\overset{M}%
{\underset{k^{\prime}=1}{\sum}}\zeta\exp\left(  -j2\pi f_{c}\left(  \tau_{\ell
k}-\tau_{\ell^{\prime}k^{\prime}}\right)  \right)  \left[  \mathbf{R}%
_{s}\right]  _{ii^{\prime}}dt\right\}  .\nonumber
\end{align}
\bigskip In matrix form,
\begin{equation}
\mathbf{V}_{c}=\left[
\begin{array}
[c]{cc}%
\frac{\partial}{\partial\mathbf{\tau}}\operatorname{Re}\left\{  \zeta\left[
\operatorname*{diag}(\mathbf{e)R}_{s}\right]  \mathbf{e}^{H}\right\}  ; &
-\frac{\partial}{\partial\mathbf{\tau}}\operatorname{Im}\left\{  \zeta\left[
\operatorname*{diag}(\mathbf{e)R}_{s}\right]  \mathbf{e}^{H}\right\}
\end{array}
\right]  .\label{e:Vc}%
\end{equation}

\emph{Orthogonal Waveforms }

Orthogonality implies that all cross elements $\int{\small s}_{k}\left(
t-\tau_{\ell k}\right)  {\small s}_{k^{\prime}}^{\ast}\left(  t-\tau
_{\ell^{\prime}k^{\prime}}\right)  dt=o$ for $\ell\neq\ell^{\prime}$ and
$k\neq k^{\prime}$. Therefore, the matrices defined by (\ref{eq:App_b4}%
)-(\ref{eq:App_b9}) take the following form:%
\begin{equation}%
\begin{array}
[c]{c}%
\left[  \mathbf{S}_{c_{or}}\right]  _{ii^{\prime}}=\left\{
\begin{array}
[c]{cc}%
4\pi^{2}\left\vert \zeta\right\vert ^{2}f_{c}^{2}f_{R_{k}} & i=i^{\prime}\\
0 & i\neq i^{\prime}%
\end{array}
\right. \\
\lbrack\mathbf{\Lambda}_{\alpha c_{or}}]_{11}=[\mathbf{\Lambda}_{\alpha_{or}%
}]_{22}=\left\{
\begin{array}
[c]{cc}%
\frac{1}{MN} & i=i^{\prime}\\
0 & i\neq i^{\prime}%
\end{array}
\right. \\
\lbrack\mathbf{\Lambda}_{\alpha c_{or}}]_{21}=[\mathbf{\Lambda}_{\alpha_{or}%
}]_{12}=0\\
\lbrack\mathbf{V}_{c_{or}}]_{i1}=2\pi\zeta^{I}f_{c}\\
\lbrack\mathbf{V}_{c_{or}}]_{i2}=-2\pi\zeta^{R}f_{c}.
\end{array}
\label{eq:App_b10}%
\end{equation}

where $f_{R_{k}}=\left(  1+\frac{\beta_{k}^{2}}{f_{c}^{2}}\right)  $. When we
invoke the narrowband assumption $\beta_{k}^{2}/f_{c}^{2}\ll1$ it follows that
$f_{R_{k}}\simeq1$.%

\section{Computation of (36)}%

\label{Section:appendixC}

The submatrix $\left[  \mathbf{C}_{CRLB_{c}}\right]  _{2\times2}$ is defined
as:
\begin{equation}
\left[  \mathbf{C}_{CRLB_{c}}\right]  _{2\times2}=\left[  \mathbf{J}\left(
\mathbf{\theta}_{c}\right)  \right]  _{2\times2}^{-1}.\label{eq:a1eq1}%
\end{equation}

For a given matrix of the form:%

\begin{equation}
\mathbf{J}\left(  \mathbf{\theta}_{c}\right)  =\left[
\begin{array}
[c]{cc}%
\mathbf{HS}_{c}\mathbf{H}^{T} & \mathbf{HV}_{c}\\
\mathbf{V}_{c}^{T}\mathbf{H}^{T} & \mathbf{\Lambda}_{{\normalsize \alpha c}}%
\end{array}
\right]  ,\label{eq:a1eq2}%
\end{equation}
where $\mathbf{\Lambda}_{{\normalsize \alpha c}}$ is a diagonal matrix of the
form $\mathbf{\Lambda}_{{\normalsize \alpha c}}=d\mathbf{I}_{2\times2}$, and
$d$ is some constant.

By definition, the value of $\left[  \mathbf{J}\left(  \mathbf{\theta}%
_{c}\right)  \right]  _{1,1}^{-1}$ is obtained by:%

\begin{equation}
\left[  \mathbf{J}\left(  \mathbf{\theta}_{c}\right)  \right]  _{1,1}%
^{-1}=\frac{\left\vert \widetilde{\mathbf{J}}\left(  \mathbf{\theta}%
_{c}\right)  _{ex\left(  1,1\right)  }\right\vert }{\left\vert \mathbf{J}%
\left(  \mathbf{\theta}_{c}\right)  \right\vert },\label{eq:AppCR1a}%
\end{equation}
\ where $\left\vert \mathbf{\cdot}\right\vert $ denotes the determinant, and
$\widetilde{\mathbf{J}}\left(  \mathbf{\theta}\right)  _{ex\left(  1,1\right)
}$\ is a submatrix, obtained by removing the first row and the first column of
the $\mathbf{J}\left(  \mathbf{\theta}_{c}\right)  $\ matrix. The determinant
of $\mathbf{J}\left(  \mathbf{\theta}_{c}\right)  $, using the property that
the determinant of a matrix does not change under linear operations, is:%

\begin{equation}
\left\vert \mathbf{J}\left(  \mathbf{\theta}_{c}\right)  \right\vert
=\left\vert
\begin{array}
[c]{cc}%
\mathbf{H\mathbf{S}}_{c}\mathbf{H}^{T}-\mathbf{V}_{c}^{T}\mathbf{H}%
^{T}\mathbf{\Lambda}_{{\normalsize \zeta}}^{-1}\mathbf{HV}_{c} & \mathbf{0}\\
\mathbf{V}_{c}^{T}\mathbf{H}^{T} & \mathbf{\Lambda}_{{\normalsize \alpha c}}%
\end{array}
\right\vert .\label{eq:AppCR4a}%
\end{equation}

This can be calculated and expressed as:%

\begin{equation}
\left\vert \mathbf{J}\left(  \mathbf{\theta}_{c}\right)  \right\vert
=\left\vert \mathbf{H\mathbf{\mathbf{S}}}_{c}\mathbf{H}^{T}-\mathbf{V}_{c}%
^{T}\mathbf{H}^{T}\mathbf{\Lambda}_{{\normalsize \zeta}}^{-1}\mathbf{HV}%
\right\vert \left\vert \mathbf{\Lambda}_{{\normalsize \alpha c}}\right\vert
.\label{eq:AppCR6a}%
\end{equation}

Repeating the same for the matrix $\widetilde{\mathbf{J}}\left(
\mathbf{\theta}_{c}\right)  _{ex\left(  1,1\right)  }$:%

\begin{equation}
\widetilde{\mathbf{J}}\left(  \mathbf{\theta}_{c}\right)  _{ex\left(
1,1\right)  }=\left[
\begin{array}
[c]{cc}%
\widetilde{\mathbf{H\mathbf{\mathbf{S}}}_{c}\mathbf{H}^{T}}_{ex\left(
1,1\right)  } & \widetilde{\mathbf{HV}_{c}}_{ex\left(  1,\right)  }\\
\widetilde{\mathbf{V}_{c}^{T}\mathbf{H}^{T}}_{ex\left(  ,1\right)  } &
\mathbf{\Lambda}_{{\normalsize \alpha c}}%
\end{array}
\right]  .\label{eq:AppCR7a}%
\end{equation}

Using the same matrix manipulation, we get:%

\begin{equation}
\left\vert \widetilde{\mathbf{J}}\left(  \mathbf{\theta}_{c}\right)
_{ex\left(  1,1\right)  }\right\vert =\left\vert \widetilde
{\mathbf{H\mathbf{S}}_{c}\mathbf{H}^{T}}-\widetilde{\mathbf{V}_{c}%
^{T}\mathbf{H}^{T}}\mathbf{\Lambda}_{{\normalsize \alpha c}}^{-1}%
\widetilde{\mathbf{HV}_{c}}\right\vert \left\vert \mathbf{\Lambda
}_{{\normalsize \alpha c}}\right\vert ,\label{eq:AppCR8}%
\end{equation}
and using terms (\ref{eq:AppCR6a}) and (\ref{eq:AppCR8}) in (\ref{eq:AppCR1a}) yields:%

\begin{equation}
\left[  \mathbf{J}\left(  \mathbf{\theta}_{c}\right)  \right]  _{1,1}%
^{-1}=\frac{\left\vert \widetilde{\mathbf{H\mathbf{S}}_{c}\mathbf{H}^{T}%
}-\widetilde{\mathbf{V}_{c}^{T}\mathbf{H}^{T}}\mathbf{\Lambda}%
_{{\normalsize \alpha c}}^{-1}\widetilde{\mathbf{HV}_{c}}\right\vert
}{\left\vert \mathbf{H\mathbf{S}}_{c}\mathbf{H}^{T}-\mathbf{V}_{c}%
^{T}\mathbf{H}^{T}\mathbf{\Lambda}_{{\normalsize \alpha c}}^{-1}%
\mathbf{HV}_{c}\right\vert }.\label{eq:AppCR9a}%
\end{equation}

By definition, this expression is identical to:%

\begin{equation}
\left[  \mathbf{J}\left(  \mathbf{\theta}_{c}\right)  \right]  _{1,1}%
^{-1}{\normalsize =}\left[  \left(  \mathbf{H\mathbf{S}}_{c}\mathbf{H}%
^{T}-\mathbf{V}_{c}^{T}\mathbf{H}^{T}\mathbf{\Lambda}_{{\normalsize \alpha c}%
}^{-1}\mathbf{HV}_{c}\right)  ^{-1}\right]  _{1,1}.\label{eq:AppCR10a}%
\end{equation}

Repeating the process for term located at $\left(  1,2\right)  $, $\left(
2,1\right)  $, and $\left(  2,2\right)  $, \ results in:%

\begin{equation}
\left[  \mathbf{C}_{C{\normalsize RLB}_{c_{or}}}\right]  _{2\times
2}{\normalsize =}\left(  \mathbf{H\mathbf{S}}_{c}\mathbf{H}^{T}-\mathbf{V}%
_{c}^{T}\mathbf{H}^{T}\mathbf{\Lambda}_{{\normalsize \alpha c}}^{-1}%
\mathbf{HV}_{c}\right)  ^{-1}.\label{eq:app_c2}%
\end{equation}
%

\section{Derivation of Covariance of Observation Noise (78)}%

\label{Section:appendixD}

For a set of received waveforms $r_{\ell}\left(  t\right)  ,$ $1\leq\ell\leq
N, $ the time delay estimates $\mu=\left[  \mu_{11},\mu_{12},...,\mu
_{MN}\right]  ^{T}$ are determined by maximizing the following statistic:%

\begin{equation}
\mu_{\ell k}=\arg\max_{v}\left[  \exp\left(  j2\pi f_{c}v\right)  \int
_{T}r_{\ell}\left(  t\right)  s_{k}^{\ast}\left(  t-v\right)  dt\right]
.\label{eq:AppTOA1}%
\end{equation}
Equivalently,
\begin{equation}
\frac{d}{dv}\left[  \exp\left(  j2\pi f_{c}v\right)  \int_{T}r_{\ell}\left(
t\right)  s_{k}^{\ast}\left(  t-v\right)  dt\right]  _{v=\mu_{\ell k}%
}=0.\label{e:tde}%
\end{equation}
The time delay estimates are expressed in (\ref{e:mu}). The properties of the
noise $\epsilon_{\ell k}$ can be computed from (\ref{e:td}), and (\ref{e:r}).
It is not difficult to show that the following relation holds:%
\begin{equation}
\left.  \frac{dg(v)}{dv}\right\vert _{v=\mu_{\ell k}}+n_{\ell k}%
=0,\label{eq:AppTOA5}%
\end{equation}
where
\begin{equation}
g(v)=\zeta\int_{T}\exp\left[  j2\pi f_{c}\left(  v-\tau_{\ell k}\right)
\right]  s_{k}\left(  t-\tau_{\ell k}\right)  s_{k}^{\ast}\left(  t-v\right)
dt,\label{eq:AppTOA4}%
\end{equation}
and%
\begin{equation}
n_{\ell k}=\int_{T}\frac{d}{dv}w_{\ell}(t)s_{k}^{\ast}\left(  t-v\right)
\exp\left(  j2\pi f_{c}v\right)  dt.\label{eq:TOAtoo}%
\end{equation}
We wish to write (\ref{eq:AppTOA5}) in the form of (\ref{e:mu}). With a few
algebraic manipulations, including expanding $g(v)$ in a Taylor series around
$\tau_{\ell k},$ and neglecting terms $o\left[  \left(  \tau_{\ell k}%
-\widehat{\tau}_{\ell k}\right)  ^{3}\right]  ,$ it can be shown that
\begin{equation}
\mu_{\ell k}=\tau_{\ell k}+\frac{n_{\ell k}}{4\pi^{2}f_{c}^{2}\left(
1+\frac{\beta_{k}^{2}}{f_{c}^{2}}\right)  \zeta}.\label{eq:AppTOA6}%
\end{equation}
Comparing this with (\ref{e:mu}), and invoking the narrowband assumption
$\beta_{k}^{2}/f_{c}^{2}\ll1$, we have for the error term
\begin{equation}
\epsilon_{\ell k}\simeq\frac{n_{\ell k}}{4\pi^{2}\zeta f_{c}^{2}%
}.\label{eq:AppTOA13a}%
\end{equation}

To find the first and second order statistics of $\epsilon_{\ell k},$ we need
the statistical characterization of $n_{\ell k}$. As previously stated, we
assume the receiver noise $w_{\ell}(t)$ is a Gaussian random process with zero
mean and autocorrelation function $\sigma_{w}^{2}\delta(\tau)$. Since $n_{\ell
k}$\ is a linear transformation of the process $w_{\ell}(t),$ since the mean
$w_{\ell}(t)$ is zero, $E\left[  n_{\ell k}\right]  =0.$ Similarly, it can be
shown that%

\begin{equation}
E\left[  n_{\ell k}{}n_{nm}^{\ast}\right]  =\left\{
\begin{array}
[c]{lc}%
0 & \forall\ell k\neq nm\\
2\pi^{2}\sigma_{w}^{2}f_{c}^{2} & \forall\ell k=nm\text{ }%
\end{array}
\right.  .\label{eq:AppTOA14}%
\end{equation}
Using these results, we finally get
\begin{align}
E\left[  \epsilon_{\ell k}\epsilon_{nm}^{\ast}\right]   &  =\frac{E\left[
n_{\ell k}{}n_{nm}\right]  }{16\pi^{4}\left\vert \zeta\right\vert ^{2}%
f_{c}^{4}}\label{eq:AppTOA15}\\
&  =\left\{
\begin{array}
[c]{lc}%
0 & \forall\ell k\neq nm\\
\frac{1}{8\pi^{2}f_{c}^{2}\left(  \left\vert \zeta\right\vert ^{2}/\sigma
_{w}^{2}\right)  } & \forall\ell k=nm\text{ }%
\end{array}
\right.  ,\nonumber
\end{align}
concluding that the covariance matrix of the terms $\epsilon_{\ell k}$ is
given by:%

\begin{equation}
\mathbf{C}_{\mathbf{\epsilon}}=\frac{1}{8\pi^{2}f_{c}^{2}\left\vert
\zeta\right\vert ^{2}/\sigma_{w}^{2}}\mathbf{I}_{MN\times MN}.
\end{equation}

%

%

%

%

%

\bigskip

\end{document}